\documentclass[aoas,MSNbibl,nameyear,rotating,dvips]{arximspdf}
\usepackage{dcolumn}
\usepackage{graphicx}
\usepackage{url,breakurl}
\doi{10.1214/14-AOAS738} %kopijuoti is PTS
\volume{8}
\issue{3}
\pubyear{2014}
\firstpage{1469}
\lastpage{1491}

\makeatletter
\newcolumntype{d}[1]{D{.}{.}{#1}}
\renewcommand{\mid}{|}

\makeatother

\begin{document}
\begin{frontmatter}

\title{Rank discriminants for predicting phenotypes from~RNA expression}
\runtitle{Rank discriminants}

\begin{aug}
\author[A]{\fnms{Bahman}~\snm{Afsari}\corref{}\thanksref{T1,M1}\ead[label=e1]{bahman@jhu.edu}},
\author[B]{\fnms{Ulisses M.}~\snm{Braga-Neto}\thanksref{T2,M2}\ead[label=e2]{ub@ieee.org}}
\and
\author[C]{\fnms{Donald}~\snm{Geman}\thanksref{T1,M1}\ead[label=e3]{geman@jhu.com}}
\runauthor{B. Afsari, U. M. Braga-Neto and D. Geman}
\affiliation{Johns Hopkins University\thanksmark{M1} and Texas A\&M
University\thanksmark{M2}}
\address[A]{B. Afsari\\
Department of Electrical\\
\quad and Computer Engineering\\
Johns Hopkins University\\
Baltimore, Maryland 21218\\
USA\\
\printead{e1}}
\address[B]{U. M. Braga-Neto\\
Department of Electrical\\
\quad and Computer Engineering\\
Texas A\&M University\\
College Station, Texas 77843\\
USA\\
\printead{e2}}%adresu isvedimo komanda gale!
\address[C]{D. Geman\\
Department of Applied Mathematics and Statistics\\
Johns Hopkins University\\
Baltimore, Maryland 21218\\
USA\\
\printead{e3}}
\end{aug}
\thankstext{T1}{Supported by NIH-NCRR Grant UL1 RR 025005.}
\thankstext{T2}{Supported by the National Science Foundation through
award CCF-0845407.}

% HISTORY:
\received{\smonth{12} \syear{2012}}
\revised{\smonth{3} \syear{2014}}

% ABSTRACT
%
\begin{abstract}
Statistical methods for analyzing large-scale biomolecular data are
commonplace in computational biology. A notable example is phenotype
prediction from gene expression data, for instance, detecting human
cancers, differentiating subtypes and predicting clinical outcomes.
Still, clinical applications remain scarce. One reason is that the
complexity of the decision rules that emerge from standard statistical
learning impedes biological understanding, in particular, any
mechanistic interpretation. Here we explore decision rules for binary
classification utilizing only the ordering of expression among several
genes; the basic building blocks are then two-gene expression
comparisons. The simplest example, just one comparison, is the \textit{TSP}
classifier, which has appeared in a variety of cancer-related
discovery studies. Decision rules based on multiple comparisons can
better accommodate class heterogeneity, and thereby increase accuracy,
and might provide a link with biological mechanism. We consider a
general framework (``rank-in-context'') for designing discriminant
functions, including a data-driven selection of the number and
identity of the genes in the support (``context''). We then
specialize to two examples: voting among several pairs and comparing
the median expression in two groups of genes. Comprehensive
experiments assess accuracy relative to other, more complex, methods,
and reinforce earlier observations that simple classifiers are
competitive.
\end{abstract}

% KEYWORDS
% Pirmas kwd is didziosios raides
%
\end{frontmatter}

%s1 #&#
\section{Introduction}\label{sec1}

Statistical methods for analyzing high-dimensional bio\-molecular data
generated with high-throughput technologies permeate the literature in
computational biology. Such analyses have uncovered a great deal of
information about biological processes, such as important mutations
and lists of ``marker genes'' associated with common diseases
[\citet{Vogelestein08}, \citet{MutationOncogenes}] and key interactions in
transcriptional regulation [\citet{BioMarkers}, \citet{Network1}]. Our interest
here is learning classifiers that can distinguish between cellular
phenotypes from mRNA transcript levels collected from cells in assayed
tissue, with a primary focus on the structure of the prediction
rules. Our work is motivated by applications to genetic diseases such
as cancer, where malignant phenotypes arise from the net effect of
interactions among multiple genes and other molecular agents within
biological networks. Statistical methods can\vadjust{\goodbreak} enhance our understanding
by detecting the presence of disease (e.g., ``tumor'' vs ``normal''),
discriminating among cancer subtypes (e.g., ``GIST'' vs ``LMS'' or
``BRCA1 mutation'' vs ``no BRCA1 mutation'') and predicting clinical
outcomes (e.g., ``poor prognosis'' vs ``good prognosis'').

Whereas the need for statistical methods in biomedicine continues to
grow, the effects on clinical practice of existing classifiers based
on gene expression are widely acknowledged to remain limited;
see \citet{ClinicalLimits1}, \citet{ClinicalLimits2}, \citet{ClinicalLimits3} and the
discussion in
\citet{EmergCompBio2013}. One barrier is the study-to-study diversity
in reported prediction accuracies and ``signatures'' (lists of
discriminating genes). Some of this variation can be attributed to
the overfitting that results from the unfavorable ratio of the sample
size to the number of potential biomarkers, that is, the infamous ``small
$n$, large $d$'' dilemma. Typically, the number of samples (chips,
profiles, patients) per class is $n={}$10--1000, whereas the number of
features (exons, transcripts, genes) is usually $d={}$1000--50,000; Table~\ref{TableClassSizes} displays the sample sizes and the numbers of features for twenty-one
publicly available data sets involving two phenotypes.

%t1 #&#
\begin{table}
\tabcolsep=0pt
\caption{The data sets:
twenty-one data sets involving two disease-related phenotypes (e.g., cancer vs normal tissue or two cancer subtypes),
illustrating the ``small $n$, large $d$'' situation.
% Shown are the samples sizes for the two classes
% and the number $d$ of features (probes on the microarray).
The more pathological phenotype is labeled as class 1 when this information is available}\label{TableClassSizes}
\begin{tabular*}{\tablewidth}{@{\extracolsep{4in minus 4in}}@{}lcccd{5.0}c@{}}
\hline
& \textbf{Study} & \textbf{Class 0 (size)} & \textbf{Class 1 (size)} & \multicolumn{1}{c}{\textbf{Probes $\bolds{d}$}} & \textbf{Reference}\\
\hline
D1&Colon & Normal (22) &Tumor (40) & 2000 & \citet{ColonDataSet}\\
D2&BRCA1 & Non-BRCA1 (93) &BRCA1 (25) & 1658 & \citet{TSTPaper}\\
D3&CNS & Classic (25) & Desmoplastic (9) & 7129 & \citet{CNSDataSet}\\
D4& DLBCL & DLBCL (58) & FL (19) & 7129 & \citet{DLBCLDataSet}\\
D5&Lung & Mesothelioma (150) & ADCS (31) & 12{,}533 & \citet{LungCancer} \\
D6&Marfan & Normal (41) & Marfan (60) & 4123 & \citet{MarfanDataSet}\\
D7&Crohn's & Normal (42) & Crohn's (59) & 22{,}283 & \citet{Crohns}\\
D8&Sarcoma & GIST (37) & LMS (31) & 43{,}931 & \citet{PricePaper}\\
D9&Squamous & Normal (22) & Head--neck (22) & 12{,}625 & \citet{SquamousDataSet}\\
D10&GCM & Normal (90) &Tumor (190) & 16{,}063 & \citet{GCMDataSet}\\
D11&Leukemia 1 & ALL (25) & AML (47) & 7129 & \citet{LeukemiaDataSet} \\
D12&Leukemia 2 & AML1 (24) & AML2 (24) & 12{,}564 & \citet{LeukemiaG}\\
D13&Leukemia 3 &ALL(710)& AML (501)& 19{,}896 & \citet{Leukemia3} \\
D14&Leukemia 4 &Normal (138)&AML (403) & 19{,}896& \citet{Leukemia4}\\
D15&Prostate 1 & Normal (50) & Tumor (52) &12{,}600 & \citet{Prostate1}\\
D16&Prostate 2 & Normal (38) & Tumor (50) & 12{,}625 & \citet{Prostate2}\\
D17&Prostate 3 & Normal (9) &Tumor (24) & 12{,}626 & \citet{Prostate3}\\
D18&Prostate 4 & Normal (25) & Primary (65) & 12{,}619 &\citet{ProstateMGDataSet}\\
D19&Prostate 5 & Primary (25) & Metastatic (65) & 12{,}558 & \citet{ProstateMGDataSet}\\
D20&Breast 1& ER-positive (61)& ER-negative(36)&16{,}278&\citet{ERGSE19783}\\
D21&Breast 2& ER-positive (127)&ER-negative (80)&9760&\citet{ERGSE22220}\\
\hline
\end{tabular*}
\end{table}

Complex decision rules are obstacles to
mature applications. The classification methods applied to biological
data were usually designed for other purposes, such as improving
statistical learning or applications to vision and speech, with little
emphasis on transparency. Specifically, the rules generated by nearly
all standard, off-the-shelf techniques applied to genomics data, such
as neural networks [\citet{NN1}, \citet{NN2}, \citet{NN3}], multiple
decision trees [\citet{DT1}, \citet{DT2}], support vector machines
[\citet{SVM1}, \citet{SVM2}], boosting [\citet{Boosting2}, \citet{Boosting1}] and linear discriminant analysis [\citet
{SCRDA07,PAM02}], usually involve nonlinear functions of hundreds or
thousands of genes, a great many parameters, and are therefore too
complex to characterize mechanistically.

In contrast, follow-up studies, for instance, independent validation or
therapeutic development, are usually based on a relatively small
number of biomarkers and usually require
an understanding of the role of the genes and gene
products in the context of molecular pathways. Ideally, the decision
rules could be interpreted mechanistically, for instance, in terms of
transcriptional regulation, and be robust with respect to parameter
settings. Consequently, what is notably missing
from the large body of work applying classification methodology to
computational genomics is a solid link with potential mechanisms, which
seem to be a necessary condition for ``translational medicine''
[\citet{EmergCompBio2013}], that is, drug development and
clinical diagnosis.

% Needless to say, accuracy is also necessary, but the accuracy of many
% of the methods mentioned above is already high enough to be of
% potential clinical value for many important phenotype distinctions.
% \textcolor{red}{Further,} it has become commonplace to follow
%methodological
% development and illustrations on real data with a discussion of the
% genes appearing in the support (``signature'') of the classifier,
% often in terms of their ``enrichment'' for specific biological process
% and molecular functions, this does not substitute for providing a
% potential mechanistic characterization of the decision rules in terms
% of biochemical interactions or specific regulatory motifs.

These translational objectives, and small-sample issues,
argue for limiting the number of parameters and introducing strong
constraints. The two principal objectives for the family of classifiers
described here are as follows:
\begin{itemize}
\item Use elementary and parameter-free building
blocks to assemble a classifier which is determined by
its support.
\item Demonstrate that such classifiers can be as discriminating as those
that emerge from the most powerful methods in statistical learning.
\end{itemize}

The building blocks we choose are two-gene comparisons, which we view
as ``biological switches'' which can be directly related to
regulatory ``motifs'' or other properties of transcriptional networks.
The decision rules are then determined by expression orderings.
However, explicitly connecting statistical classification and
molecular mechanism for particular diseases is a major challenge and
is well beyond the scope of this paper; by our construction we are
anticipating our longer-term goal of incorporating mechanism by
delineating candidate motifs using prior biological knowledge. Some
comments on the relationship between comparisons and regulation appear
in the concluding section.

To meet our second objective, we measure the performance of our\break
comparison-based classifiers relative to two popular alternatives,
namely, support vector machines and \emph{PAM} [\citet{PAM02}], a
variant of
linear discriminant \mbox{analysis}. The ``metric'' chosen is the estimated
error in multiple runs of tenfold cross-validation for each of the
twenty-one real data sets in Table~\ref{TableClassSizes}. (Computational cost is not
considered; applying any of our comparison-based decision rules to a
new sample is virtually instantaneous.) Whereas a comprehensive
simulation study could be conducted,
for example, along the lines of those in \citet{guo2005regularized}, \citet{SVMNonConv} and \citet{fan2008high} based on
Gaussian models of
microarray data, rather our intention is different: show that even when the
number of parameters is small, in fact, the decision rule is determined
by the support, the accuracy measured by cross-validation on real data
is no
worse than with currently available classifiers.

\begin{figure}

\includegraphics{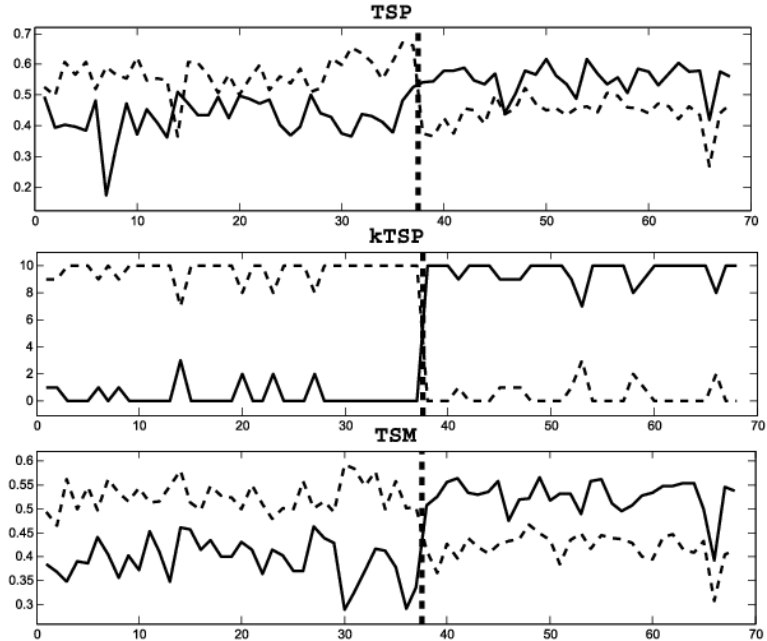}

\caption{Results of three rank-based
classifiers for differentiating two cancer subtypes, GIST and LMS.
The training set consists of 37 GIST samples and 31 LMS samples
(separated by the vertical dashed line); each sample provides
measurements for 43,931 transcripts.
\emph{TSP}: expression values for the two genes selected by the \emph{TSP} algorithm.
\emph{KTSP}: the number of votes for each class among the $K=10$ pairs of genes selected by the \emph{KTSP} algorithm.
\emph{TSM}: median expressions of two sets of genes selected by the \emph{TSM} algorithm.}\label{FigRCSarcoma}
\end{figure}

More precisely, all the classifiers studied in this paper are based on a
general \textit{rank discriminant} $g(\mathbf{X}; \Theta)$, a real-valued
function on the ranks of $\mathbf{X}$ over a (possibly ordered)
subset of genes $\Theta$, called
the \textit{context} of the classifier. We are searching for characteristic
perturbations in this ordering from one phenotype to another. The
\textit{TSP} classifier is the simplest example
(see Section~\ref{sec2}), and the decision rule is illustrated in
Figure~\ref{FigRCSarcoma}. This data set has expression profiles
for two kinds of gastrointestinal cancer (gastrointestinal
stromal-GIST, leiomyosarcoma-LMS) which are difficult to distinguish
clinically but require very different treatments
[\citet{PricePaper}]. Each point on the $x$-axis corresponds to a sample,
and the vertical dashed line separates the two phenotypes. The $y$-axis
represents expression; as seen, the ``reversal'' of the ordering of
the expressions of the two genes identifies the phenotype except in
two samples.

%f1 #&#

Evidently, a great deal of information may be lost by converting
to ranks, particularly if
the expression values are high resolution. But there are technical
advantages to basing prediction on ranks, including reducing
study-to-study variations due to data normalization and
preprocessing. Rank-based methods are evidently
invariant to general monotone transformations of the original
expression values, such as the widely-used quantile normalization
[\citet{QuantNormalization04}]. Thus, methods based on ranks can combine
inter-study microarray data without the need to
perform data normalization, thereby increasing sample size.

However, our principal motivation is complexity reduction: severely
limiting the number of variables and parameters, and in fact
introducing what we call \textit{rank-in-context} (RIC) discriminants
which depend on the training data only through the context. The
classifier $f$ is then defined by thresholding $g$. This implies
that, given a context $\Theta$, the RIC classifier corresponds to a
\textit{fixed} decision boundary, in the sense that it does not depend on
the training data. This sufficiency property helps to reduce variance
by rendering the classifiers relatively insensitive to small
disturbances to the ranks of the training data and is therefore
especially suitable to small-sample settings. Naturally, the
performance critically depends on the appropriate choice of $\Theta$.
We propose a simple yet powerful procedure to select $\Theta$ from the
training data, partly inspired by the principle of analysis of
variance and involving the sample means and sample variances of the
empirical distribution of~$g$ under the two classes. In particular,
we do not base the choice directly on minimizing error.

We consider two examples of the general framework. The first is a
new method for learning the context of \textit{KTSP}, a previous extension of
\textit{TSP} to a variable number of pairs.
The decision rule of the \textit{KTSP} classifier is
the majority vote among the top $k$ pairs of genes, illustrated in
Figure~\ref{FigRCSarcoma} for $k=10$ for the same data set as above. In
previous statistical and applied work
[\citet{kTSPPaper}], the parameter $K$ (the number of
comparisons) was
determined by an inner loop of cross-validation, which is subject to
overfitting with small samples. We also propose comparing the median
of expression between two sets of genes; this \textit{Top-Scoring Median}
(\textit{TSM}) rule is also illustrated in Figure~\ref{FigRCSarcoma}.
As can
be seen, the difference of the medians generally has a
larger ``margin'' than in the special case of singleton sets, that is,
\textit{TSP}.
%The third method is based on a natural distance between
%orderings - the minimum number of adjacent swaps needed to transform
%one permutation to another (equivalent to counting the number of pairs
%with divergent ordering). Unlike the others, this classifier does not
%have the RIC property. Given the context, the label of a new
%sample is determined by the average swap distance to the training
%samples in
%the two classes based on the order statistics. The decision rule is
%depicted for the stomach cancer dataset in Figure
A summary of all the methods is given in Table~\ref{TableSummary}.

%t2 #&#
\begin{sidewaystable}%[ht]
\tablewidth=\textwidth
\tabcolsep=0pt
\caption{Summary of rank discriminants.
First column: the rank-based classifiers considered in this paper.
Second column: the structure of the context $\Theta_k$, the genes appearing in the classifier; for \emph{kTSP} and \emph{TSM}$,\Theta_k$ contains $2k$ genes.
Third column: the form of the rank discriminant; the classifier is $f(X)=I(g(X)>0)$.
Fourth column: the selection of the context from training data.
For a fixed $k$ we select $\Theta_k$ to maximize $\hat{\delta}$, and then choose $k$ to maximize $\hat{\delta}$ normalized by $\hat{\sigma}$}\label{TableSummary}
\begin{tabular*}{\tablewidth}{@{\extracolsep{\fill}}@{}lccc@{}}
\hline
& \textbf{Parameters} & \textbf{Discriminant} & \textbf{Parameter selection}\\
\hline
General & $(\Theta_k,k)$& $g(X;\Theta_k)$& \\
 & $\Theta_k \subset\{1,\ldots,d\}$ & $\hat{\delta}(\Theta_k ) = \widehat{E}(g(X;\Theta_k ) | Y = 1 )  - \widehat{E}(g(X;\Theta_k ) | Y = 0)$ & $\Theta^*_k =\arg\max_{\Theta_k}\hat{\delta}(\Theta_k)$ \\% s.t. $\Theta^*_k
&&$\hat{\sigma}(\Theta_k)=\sqrt{\widehat{\operatorname
{Var}}(g|Y=0)+\widehat{\operatorname{Var}}(g|Y=1)}$&$k^*=\arg\max
_{k}\frac{\hat{\delta}(\Theta^*_k)}{\hat{\sigma}(\Theta^*_k)}$
\\
\hline
\multicolumn{4}{@{}c@{}}{Examples}\\
\hline
TSP & $\Theta=(i,j)$&$g_{\mathrm{TSP}}=I(X_i<X_j)-\frac{1}{2}$&$\Theta^*=\arg
\max_{(i,j)\in\Theta}\hat{s}_{ij}$\\
& &$\hat{s}_{ij} =  P(X_i < X_j | Y =  1 ) -   P(
X_i  < X_j| Y = 0 ) $&
\\[9pt]
KTSP&$\Theta_k=\{i_1,j_1,\ldots,i_k,j_k\}$ &$g_{\mathrm{KTSP}}=\sum
_{r=1}^k[I(X_{i_r}<X_{j_r})-\frac{1}{2}]$&$\Theta^*_k=\arg\max
_{\Theta_k}\sum_{r=1}^k\hat{s}_{i_rj_r}$
%& $k=|\Theta_k|$(*) &$Z_{ij}=I\{X_i<X_j\}$&\\
\\[9pt]
TSM&$\Theta_k=G^+_k\cup G^-_k$&$g_{\mathrm{TSM}}=\operatorname{med}
_{j \in G^+_k}R_j-\operatorname{med}_{i \in G^-_k}R_i$& \\
&$G^-_k=\{i_1,\ldots,i_k\}$&$R_i$: rank of gene $i$ in $G^+_k\cup
G^-_k$&$\Theta^*_k\approx\arg\max_{\Theta_k}\sum
_{i\in G^-_k,j\in G^+_k} \hat{s}_{ij}$\\
&$G^+_k=\{j_1,\ldots,j_k\}$& &
\\
% TSN&$\Theta_k=G^+_k\cup G^-_k$&$g_{RS}=\frac{1}{k}\sum_{j
% \hline
% SWP&$\Theta_k= G_k$&$g_{SWP}(X;\Theta_k) $ where &\\
% & $G_k=\{i_1,\ldots,i_k\}$&$g_{SWP}(x;\Theta_k)=\widehat{E}[D(x,X';
%_{i,j\in\Theta_k} \widehat{s}_{ij}^2$\\
% &&$-\widehat{E}[D(x,X';\Theta_k)|Y'=1]$&\\
\hline
\end{tabular*}
\end{sidewaystable}

After reviewing related work in the following section, in Section~\ref{sec3} we
present the classification scenario, propose our general statistical
framework and focus on two examples: \emph{KTSP} and \emph{TSM}. The
experimental results are in Section~\ref{sec4}, where comparisons are drawn,
and we conclude with some discussion about the underlying biology in
Section~\ref{sec5}.

%s2 #&#
\section{Previous and related work}\label{sec2}

Our work builds on previous studies analyzing transcriptomic data
solely based on the \emph{relative expression} among a small number of
transcripts.
The simplest
example, the Top-Scoring Pair (\emph{TSP}) classifier, was introduced
in \citet{Geman04} and is based on two genes. Various extensions and
illustrations appeared in \citet{uTSPPaper}, \citet{TSTPaper} and
\citet{kTSPPaper}. Applications to phenotype classification include
differentiating between stomach cancers [\citet{PricePaper}],
predicting treatment response in breast cancer
[\citet{BreastWeichselbaum}] and acute myeloid leukemia
[\citet{BloodTSP}], detecting BRCA1 mutations [\citet{TSTPaper}],
grading prostate cancers [\citet{ProstateTSP}] and separating diverse
human pathologies assayed through blood-borne leukocytes
[\citet{Edelman09}].

In \citet{Geman04} and subsequent papers about \emph{TSP}, the
discriminating power of each pair of genes $i,j$ was measured by
the absolute difference between the probabilities of the event that gene
$i$ is expressed more than gene $j$ in the two classes. These
probabilities were estimated from training data and (binary)
classification resulted from voting among all top-scoring pairs. In
\citet{uTSPPaper} a secondary score was introduced which provides a
\emph{unique} top-scoring pair. In addition, voting was
extended to the $k$ highest-scoring pairs of genes. The motivation
for this \emph{KTSP} classifier and other extensions
[\citet{kTSPPaper}, \citet{kTSPProtonomics07}, \citet
{TSPGPaper07}] is that
more genes may be needed to detect cancer pathogenesis, especially if
the principle objective is to characterize as well as recognize the process.
Finally, in a precursor to the work
here [\citet{TSPGPaper07}], the two genes in \emph{TSP} were
replaced by two
equally-sized \emph{sets} of genes and the average ranks were compared.
Since the direct extension of \emph{TSP} score maximization was
computationally impossible, and likely to badly overfit the data, the
sets were selected by splitting top-scoring pairs and repeated random
sampling. Although ad hoc, this process further
demonstrated the discriminating power of rank statistics for
microarray data.

Finally, there is some related work about
ratios of concentrations (which are natural
in chemical terms)
for diagnosis and prognosis. That work is not rank-based but retains
invariance to scaling. \citet{LeukemiaDataSet} distinguished between
malignant pleural mesothelioma (MPM) and adenocarcinoma (ADCA) of the
lung by combining multiple ratios into a single diagnostic tool,
and \citet{Ma04} found that a two-gene expression ratio derived
from a
genome-wide, oligonucleotide microarray analysis of estrogen receptor
(ER)-positive, invasive breast cancers predicts tumor relapse and
survival in patients treated with tamoxifen, which is crucial for
early-stage breast cancer management.
%This observation was also
%confirmed in real-time quantitative PCR analyses. Specifically, the
%expression ratio of HOXB13 over IL-17BR outperformed existing
%biomarkers for prognosis of breast cancer, such as patient age, tumor
%size, grade, and lymph node status.

%s3 #&#
\section{Rank-in-context classification}\label{sec3}

In this section we introduce a general framework for rank-based
classifiers using comparisons among a limited number of gene
expressions, called the \emph{context}. In addition, we describe a
general method to select the context, which is inspired by the
analysis of variance paradigm of classical statistics. These
classifiers have the RIC property that they depend on the sample
training data
solely through the context selection; in other words, given the
context, the classifiers have a fixed decision boundary and do not
depend on any further learning from
the training data. For example, as will be seen in later
sections, the \emph{Top-Scoring Pair} (\emph{TSP}) classifier is RIC.
Once a
pair of genes (i.e., the context) is specified, the \emph{TSP} decision
boundary is fixed and corresponds to a \mbox{45-}degree line going through
the origin in the feature space defined by the two genes. This property
confers to RIC classifiers a \emph{minimal-training} property, which
makes them insensitive to small disturbances to the ranks of the
training data, reducing variance and overfitting, and rendering them
especially suitable to the $n \ll d$ settings illustrated
in Table~\ref{TableClassSizes}. We will
demonstrate the general RIC framework with two specific examples,
namely, the previously introduced \emph{KTSP} classifier based on
majority voting among comparisons [\citet{kTSPPaper}], as well as
a new
classifier based on the comparison of the medians, the \emph{Top-Scoring Medians} (\emph{TSM}) classifier.

%s3.1 #&#
\subsection{RIC discriminant}\label{sec3.1}\label{Sec-rankdis}

Let $\mathbf{X}= (X_1, X_2, \ldots, X_d)$ denote the expression
values of
$d$ genes on an expression microarray. Our objective is to use $\mathbf{X}$
to distinguish between two conditions or phenotypes for the cells in
the assayed tissue, denoted $Y=0$ and $Y=1$.
A classifier $f$ associates a label $f(\mathbf{X}) \in\{0,1\}$ with a given
expression vector $\mathbf{X}$. Practical classifiers are inferred from
training data, consisting of i.i.d. pairs $S_n =
\{(\mathbf{X}^{(1)},Y^{(1)}),\ldots, (\mathbf{X}^{(n)},Y^{(n)})\}$.

The classifiers we consider in this paper are all defined in terms of
a general \emph{rank-in-context discriminant} $g(\mathbf{X}; \Theta
(S_n))$, which is
defined as a real-valued function on the ranks of $\mathbf{X}$
over a subset of genes $\Theta(S_n) \subset\{1,\ldots,d\}$, which is
determined by the training data $S_n$ and is called the \emph{context}
of the
classifier (the order of indices in the context may matter).
The corresponding \emph{RIC classifier} $f$ is defined by
%
%e1 #&#
\begin{equation}
f\bigl(\mathbf{X}; \Theta(S_n)\bigr) = {\mathbf I}\bigl(g\bigl(
\mathbf{X}; \Theta(S_n)\bigr) > t\bigr) = \cases{1, &\quad$g\bigl(
\mathbf{X}; \Theta(S_n)\bigr) > t$,
\vspace*{3pt}\cr
0, &\quad otherwise,}
\label{eq-cl}
\end{equation}
where ${\mathbf I}(E)$ denotes the indicator variable of event $E$. The
threshold parameter~$t$ can be adjusted to achieve a desired
specificity and sensitivity (see Section~\ref{Sec-errrt} below);
otherwise, one usually sets $t=0$. For simplicity we will write
$\Theta$ instead of $\Theta(S_n)$, with the implicit
understanding that in RIC classification $\Theta$ is selected from the
training data $S_n$.

%%% KTSP

We will consider two families of RIC classifiers. The
first example is the \emph{$k$-Top Scoring Pairs} (\emph{KTSP})
classifier, which is a majority-voting rule among $k$ pairs of genes
[\citet{kTSPPaper}]; \emph{KTSP} was the winning entry of the
International Conference in Machine Learning and Applications (ICMLA)
2008 challenge for micro-array classification [\citet{ICMLA08}].
Here, the context is partitioned into a set of gene pairs
$\Theta=\{(i_1,j_1),\ldots,(i_k,j_k)\}$, where $k$ is a positive odd
integer, in such a way that all pairs are disjoint, that is, all $2k$
genes are distinct. The RIC discriminant is given by
%
%e2 #&#
\begin{equation}
g_{\mathrm{KTSP}}\bigl(\mathbf{X};(i_1,j_1),\ldots,(i_k,j_k)\bigr) = \sum_{r=1}^k
\biggl[{\mathbf I}(X_{i_r}<X_{j_r})- \frac{1}{2} \biggr].
\label{eqKTSPStatistic}
\end{equation}
This \emph{KTSP} RIC discriminant simply counts positive and negative
``votes'' in favor of
ascending or descending ranks, respectively. The \emph{KTSP} classifier
is given by~(\ref{eq-cl}), with $t=0$, which yields
%
%e3 #&#
\begin{equation}
f_{\mathrm{KTSP}}\bigl(\mathbf{X};(i_1,j_1),\ldots,(i_k,j_k)\bigr) = {\mathbf I} \Biggl( \sum
_{r=1}^k {\mathbf I}(X_{i_r}<X_{j_r})>
\frac{k}{2} \Biggr).
\end{equation}
The \emph{KTSP} classifier is thus a majority-voting
rule: it assigns label 1 to the expression profile if the number of
ascending ranks exceeds the number of descending ranks in the
context. The choice of odd $k$ avoids the possibility of a tie in the
vote. If $k=1$, then the \emph{KTSP} classifier reduces to $f_{\mathrm{TSP}}(\mathbf{X};
(i,j)) = {\mathbf I}(X_i<X_j)$, the \emph{Top-Scoring
Pair} (\emph{TSP}) classifier [\citet{Geman04}].

%%% TSM

The second example of an RIC classifier we propose is
the \emph{Top Scoring Median} (\emph{TSM}) classifier, which
compares the median rank of two sets of genes.
The median rank has the advantage that for
any individual sample the median is the value of one of the
genes. Hence, in this sense, a comparison of medians for a given sample
is equivalent to the comparison of two-gene expressions, as in
the \emph{TSP} decision rule.
%In the same way that \emph{TSP} can be mechanistically
%interpreted, so can \emph{TSM}. }
Here, the context is partitioned into two sets
of genes, $\Theta=\{G_k^+,G_k^-\}$, such that $|G_k^+| = |G_k^-| = k$,
where $k$ is again a positive odd integer, and $G_k^+$ and $G_k^-$ are disjoint,
that is, all $2k$ genes are distinct. Let $R_i$ be the rank of $X_i$ in
the context $\Theta= G_k^+ \cup G_k^-$, such that $R_i=j$ if
$X_i$ is the $j$th smallest value among the gene expression values
indexed by $\Theta$. The RIC discriminant is given by
%
%e4 #&#
\begin{equation}
g_{\mathrm{TSM}}\bigl(\mathbf{X};G_k^+,G_k^-\bigr) =
\operatorname{med}\limits
_{j \in G_k^+}R_j - \operatorname{med}\limits
_{i \in
G_k^-}R_i, \label{eqTSMStatistic}
\end{equation}
where ``$\operatorname{med}$'' denotes the median operator. The \emph{TSM} classifier
is then given by (\ref{eq-cl}), with $t=0$, which yields
%
%e5 #&#
\begin{equation}
f_{\mathrm{TSM}}\bigl(\mathbf{X};G_k^+,G_k^-\bigr) = {
\mathbf I} \Bigl(\operatorname {med}\limits
_{j \in G_k^+}R_j > \operatorname{med}
\limits_{i \in
G_k^-}R_i
\Bigr).
\end{equation}
Therefore, the \emph{TSM} classifier outputs~1 if the median of ranks
in $G_k^+$ exceeds the median of ranks in $G_k^-$, and 0 otherwise.
Notice that this is equivalent to comparing the medians of the raw
expression values directly. We remark that an obvious variation would
be to compare the average rank rather than the median rank, which
corresponds to the ``TSPG'' approach defined in
\citet{TSPGPaper07}, except that in that study, the context for
TSPG was selected by splitting a fixed number of TSPs. We observed
that the performances of the mean-rank and median-rank classifiers are
similar, with a slight superiority of the median-rank (data not shown).

%%%%%

%s3.2 #&#
\subsection{Criterion for context selection}\label{sec3.2}\label{Sec-CS}

The performance of RIC classifiers critically depends on the
appropriate choice of the context $\Theta\subset\{1,\ldots,d\}$. We
propose a simple yet powerful procedure to select
$\Theta$ from the training data $S_n$. To motivate the
proposed criterion, first note that a necessary condition for the
context $\Theta$ to yield a good classifier is that the discriminant
$g(\mathbf{X}; \Theta)$ has sufficiently distinct distributions under $Y=1$
and $Y=0$. This can be expressed by requiring that the difference
between the expected values of $g(\mathbf{X}; \Theta)$ \emph{between} the
populations, namely,
%
%e6 #&#
\begin{equation}
\delta(\Theta) = E\bigl[g(\mathbf{X}; \Theta) \mid Y=1,S_n\bigr] - E
\bigl[g(\mathbf{X}; \Theta) \mid Y=0,S_n\bigr], \label{eqE}
\end{equation}
be maximized. Notice that this maximization is with
respect to $\Theta$ alone; $g$ is fixed and chosen {a priori}.
In practice, one employs the maximum-likelihood empirical criterion
%
%e7 #&#
\begin{equation}
\hat{\delta}(\Theta) = \widehat{E}\bigl[g(\mathbf{X}; \Theta) \mid
Y=1,S_n\bigr]- \widehat{E}\bigl[g(\mathbf{X}; \Theta) \mid
Y=0,S_n\bigr], \label{eqEs}
\end{equation}
where
%
%e8 #&#
\begin{equation}
\widehat{E}\bigl[g(\mathbf{X}; \Theta) \mid Y=c,S_n\bigr] =
\frac{\sum_{i=1}^n
g(\mathbf{X}^{(i)}; \Theta) {\mathbf I}(Y^{(i)} = c)}{\sum_{i=1}^n {\mathbf
I}(Y^{(i)} = c)}, \label{eq-Ehat}
\end{equation}
for $c=0,1$.

%%% KTSP

In the case of \emph{KTSP}, the criterion in (\ref{eqE}) becomes
%
%e9 #&#
\begin{equation}
\delta_{\mathrm{KTSP}}\bigl((i_1,j_1),\ldots,(i_k,j_k)\bigr) = \sum_{r=1}^k
s_{i_rj_r}, \label{eqEKTSP}
\end{equation}
where the \emph{pairwise score} $s_{ij}$ for the pair of genes $(i,j)$
is defined as
%
%e10 #&#
\begin{equation}
s_{ij} = P(X_i<X_j \mid Y=1)-P(X_i<X_j
\mid Y=0). \label{eq-pairsc}
\end{equation}
Notice that if the pair of random variables $(X_i,X_j)$ has a
continuous distribution, so that $P(X_i = X_j) = 0$, then
$s_{ij} = -s_{ji}$. In this case
$X_i < X_j$ can be replaced by $X_i \leq X_j$ in $s_{ij}$ in (\ref{eq-pairsc}).

The empirical criterion $\hat{\delta}_{\mathrm{KTSP}}((i_1,j_1),\ldots,(i_k,j_k))$
[cf. equation~(\ref{eqEs})] is obtained by substituting in (\ref
{eqEKTSP}) the
\emph{empirical pairwise scores}% $\hat{s}_{ij}$ is given by
%
%e11 #&#
\begin{equation}\label{eq-hatsc}
\hat{s}_{ij}  = \widehat{P}(X_i<X_j
\mid Y=1)-\widehat{P}(X_i<X_j \mid Y=0).
% & =  \frac{\sum_{m=1}^n {\mathbf I}\left(\uX^{(m)}_i<\uX^{(m)}_j\right){
% & \quad-    \frac{\sum_{m=1}^n {\mathbf I}\left(\uX^{(m)}_i<\uX^{(m)}_j
\end{equation}
Here the empirical probabilities are defined by $\widehat{P}(X_i<X_j \mid
Y=c) = \widehat{E}[ {\mathbf I}(X_i<X_j) \mid Y=c ]$, for $c = 0,1$, where
the operator $\widehat{E}$ is defined in~(\ref{eq-Ehat}).

%%% TSM

For \emph{TSM}, the criterion in (\ref{eqE}) is
given~by
%
%e12 #&#
\begin{eqnarray}\label{eqETSM}
&& \delta_{\mathrm{TSM}}\bigl(G_k^+,G_k^-
\bigr)
\nonumber\\[-8pt]\\[-8pt]
&&\qquad =  E \Bigl[\operatorname{med}\limits
_{j \in G_k^+}R_j - \operatorname{med}\limits
_{i \in G_k^-}R_i
\big| Y=1 \Bigr] - E \Bigl[\operatorname{med}\limits
_{j \in G_k^+}R_j -
\operatorname {med}\limits
_{i \in G_k^-}R_i \big| Y=0 \Bigr].\nonumber
\end{eqnarray}
Proposition S1 in Supplement A [\citet{SuppAAOAS}] shows that,
under some assumptions,
%
%e13 #&#
\begin{equation}
\delta_{\mathrm{TSM}}\bigl(G_k^+,G_k^-\bigr) =
\frac{2}{k} \sum
_{i
\in G^-_k,j \in G^+_k} s_{ij},
\label{eq-TSMavg-Main}
\end{equation}
where $s_{ij}$ is defined in (\ref{eq-pairsc}).

%The following result relates $\delta_{\mathrm{TSM}}(G_k^+,G_k^-)$ to the
%pairwise scores $s_{ij}$ defined in (\ref{eq-pairsc}). For notation
%simplicity, in the following $m^+$ and $m^-$ denote the indices of the
%genes that achieve the median rank in $G^+$ and $G^-$, respectively.

%Assume that the profile vector $\uX$ has a probability density,
%which follows the following (mild) distributional smoothness conditions
%$\{X_i < X_j\}$ is conditionally independent from $\{m^-=i\}$ and $
%given $Y$.
%$Y$.
%$\end{enumerate}
%Then the criterion (\ref{eqETSM}) can be written as
% \delta_{\mathrm{TSM}}(G_k^+,G_k^-)  =  \frac{2}{k}  \sum\limits_{i \in
%G^-_k,j \in G^+_k} s_{ij},
%where $s_{ij}$ is the pairwise score defined in (\ref{eq-pairsc}).

The difference between the two criteria (\ref{eqEKTSP})
for \emph{KTSP} and
(\ref{eq-TSMavg-Main}) for \emph{TSM} for selecting the context is that the former
involves scores for $k$ expression
comparisons and the latter involves $k^2$ comparisons
since each gene $i \in G^-_k$ is paired with each gene $j \in G^+_k$.
Moreover, using the estimated solution to maximizing (\ref{eqEKTSP})
(see below) to construct $G^-_k$ and $G^+_k$ by putting the first gene from
each pair into one and the second gene from each pair into the
other does not work as well in maximizing~(\ref{eq-TSMavg-Main}) as the algorithms described below.

The distributional smoothness conditions Proposition S1
are justified if $k$ is not too large (see Supplement A [\citet{SuppAAOAS}]). %Section
Finally, the empirical criterion $\hat{\delta}_{\mathrm{TSM}}(G_k^+,G_k^-)$ can be
calculated by substituting in (\ref{eq-TSMavg-Main}) the \emph{empirical
pairwise scores} $\hat{s}_{ij}$ defined in (\ref{eq-hatsc}).

%%%%%%%%

% Other methods for TSM: For example, filtering directly to the top $k$
%up-expressed and top
% $k$ down-expressed genes is not effective. Nor is it effective to
% choose the context as in \citet{TSPGPaper07}, by simply splitting
% top-scoring pairs.

%s3.3 #&#
\subsection{Maximization of the criterion}\label{sec3.3}

Maximization of (\ref{eqE})
or (\ref{eqEs}) works well as long as
the \emph{size} of the context $|\Theta|$, that is, the number of context
genes, is kept fixed, because the criterion tends to be
monotonically increasing with $|\Theta|$, which complicates
selection. We address this problem
by proposing a modified criterion, which is partly inspired by
the principle of analysis of variance in classical statistics.
This modified criterion penalizes the addition of
more genes to the context by requiring that the variance of $g(\mathbf{X};
\Theta)$ \emph{within} the populations be minimized. The latter is
given by
%
%e14 #&#
\begin{equation}
\hat{\sigma}(\Theta) = \sqrt{\widehat{\operatorname {Var}}\bigl(g(\mathbf{X};
\Theta)\mid Y=0,S_n\bigr)+\widehat{\operatorname{Var}}\bigl(g(
\mathbf{X}; \Theta) \mid Y=1,S_n\bigr)}, \label{eq-rhos}
\end{equation}
where $\widehat{\operatorname{Var}}$ is the maximum-likelihood
estimator of the
variance,
\begin{eqnarray*}
&& \widehat{\operatorname{Var}}\bigl(g(\mathbf{X};
\Theta) \mid Y=c,S_n\bigr)
\\
&&\qquad = \frac{\sum_{i=1}^n
(g(\mathbf{X}^{(i)}; \Theta) - \widehat{E}[g(\mathbf{X}; \Theta) \mid
Y=c,S_n])^2 {\mathbf I}(Y^{(i)} = c)}{\sum_{i=1}^n {\mathbf I}(Y^{(i)} = c)},
\end{eqnarray*}
for $c=0,1$. The modified criterion to be maximized is
%
%e15 #&#
\begin{equation}
\hat{\tau}(\Theta) = \frac{\hat{\delta}(\Theta)}{\hat{\sigma}(\Theta)}. \label{eqttests}
\end{equation}
The statistic $\hat{\tau}(\Theta)$ resembles
the Welch two-sample $t$-test statistic of classical hypothesis testing
[\citet{CaseBerg02}].\vadjust{\goodbreak}

Direct maximization of (\ref{eqEs}) or (\ref{eqttests}) is in
general a hard computational problem for the numbers of genes
typically encountered in expression data. We propose instead a
greedy procedure. Assuming that a
predefined range of values $\Omega $ for the context size $|\Theta|$ is
given, the procedure is as follows:
\begin{longlist}[(2)]
\item[(1)] For each value of
$k \in\Omega $, an optimal context $\Theta^*_k$ is chosen that maximizes~(\ref{eqEs}) among all
contexts $\Theta_k$ containing $k$ genes:
% \beq
% \Theta^*_k  =  \arg\max_{|\Theta| = k}    \hat{E}(g(\uX;
% \Theta) \mid Y=0).
% \label{eq-Tk}
% \eeq
\[
\Theta^*_k = \arg\max
_{|\Theta| = k} \hat{\delta } (\Theta).
\]

\item[(2)] An optimal value $k^*$ is chosen that maximizes
(\ref{eqttests}) among all contexts $\{\Theta^*_k \mid k \in
\Omega \}$
obtained in the previous step:
\[
k^* = \arg\max
_{k \in\Omega } \hat{\tau }\bigl(\Theta^*_k\bigr).
\]
\end{longlist}

%%% KTSP

For \emph{KTSP}, the maximization in step (1) of the previous context
selection procedure becomes
%
%e16 #&#
\begin{eqnarray}\label{eq-TkKTSP}
\bigl\{\bigl({i_1^*,j_1^*}
\bigr),\ldots,\bigl(i_k^*,j_k^*\bigr)\bigr\} & =& \arg \max
_{\{({i_1,j_1}),\ldots,(i_k,j_k)\}}
\hat{\delta}_{\mathrm{KTSP}} \bigl((i_1,j_1),\ldots,(i_k,j_k)\bigr)
\nonumber\\[-8pt]\\[-8pt]
& =& \arg \max
_{\{({i_1,j_1}),\ldots,(i_k,j_k)\}} \sum_{r=1}^k
\hat{s}_{i_r j_r}.\nonumber
\end{eqnarray}

%But this maximization can clearly be performed iteratively:
We propose a greedy approach to this maximization problem: initialize
the list with the top-scoring pair of genes, then keep adding pairs to
the list whose genes have not appeared so far [ties are broken by the
secondary score proposed in \citet{uTSPPaper}]. This process is
repeated until $k$ pairs are chosen and corresponds essentially to
the same method that was proposed, for fixed $k$, in the original
paper on \emph{KTSP} [\citet{kTSPPaper}]. Thus, the
previously proposed heuristic has a justification in terms of
maximizing the separation between the rank discriminant
(\ref{eqKTSPStatistic}) across the classes.

To obtain the optimal value $k^*$, one applies step (2) of the context
selection procedure, with a range of values $k \in\Omega=
\{3,5,\ldots,K\}$, for odd $K$ ($k = 1$ can be added if 1-\textit{TSP} is
considered). Note that here
%
%e17 #&#
\begin{eqnarray}
&& \hat{\sigma}_{\mathrm{KTSP}} (\Theta)
\nonumber\\[8pt]\\[-20pt]
&&\qquad  = \sqrt{\widehat {
\operatorname{Var}} \Biggl(\sum_{r=1}^k
\bigl[{\mathbf I}(X_{i_r^*}<X_{j_r^*})\bigr] \bigg| Y = 0 \Biggr)+
\widehat{\operatorname{Var}} \Biggl(\sum_{r=1}^k
\bigl[{\mathbf I}(X_{i_r^*}<X_{j_r^*})\bigr] \bigg| Y = 1 \Biggr)}.\nonumber\hspace*{-18pt}
\end{eqnarray}
Therefore, the optimal value of $k$ is selected by
%
%e18 #&#
\begin{equation}
k^* = \arg\max
_{k = 3,5,\ldots,K} \hat{\tau}_{\mathrm{KTSP}} \bigl(\bigl(i_1^*,j_1^*
\bigr),\ldots,\bigl(i_k^*,j_k^*\bigr)\bigr),
\end{equation}
where
%
%e19 #&#
\begin{eqnarray}\label{eq-htauTSP}
&& \hat{\tau}_{\mathrm{KTSP}}\bigl(
\bigl(i_1^*,j_1^*\bigr),\ldots,\bigl(i_k^*,j_k^*
\bigr)\bigr)\nonumber\hspace*{-6pt}
\\
&&\qquad = \frac{\hat{\delta}_{\mathrm{KTSP}}((i_1^*,j_1^*),\ldots,(i_k^*,j_k^*))}{\hat{\sigma}_{\mathrm{KTSP}}((i_1^*,j_1^*),\ldots,(i_k^*,j_k^*))}\hspace*{-6pt}
\\
&&\qquad  = \frac{\sum_{r=1}^k \hat{s}_{i_r^* j_r^*}}{
\sqrt{\widehat{\operatorname{Var}} (\sum_{r=1}^k [{\mathbf
I}(X_{i_r^*}<X_{j_r^*})] \mid Y = 0 )+\widehat{\operatorname{Var}}
 (\sum_{r=1}^k [{\mathbf
I}(X_{i_r^*}<X_{j_r^*})] \mid Y = 1 )}}.\nonumber\hspace*{-6pt}
\end{eqnarray}
Finally, the optimal context is then given by
$\Theta^*=\{(i_1^*,j_1^*),\ldots,(i^*_{k^*},j^*_{k^*})\}$.

%%% TSM

For \emph{TSM}, the maximization in step (1) of the context selection procedure
can be written~as
%
%e20 #&#
\begin{eqnarray}\label{eq-TkTSM}
\qquad \bigl(G_k^{+,*},G_k^{-,*}
\bigr) & =& \arg \max
_{(G_k^+,G_k^-)} \hat{\delta}_{\mathrm{TSM}}\bigl(G_k^+,G_k^-
\bigr)
 = \arg \max
_{(G_k^+,G_k^-)} \sum
_{i \in G^-_k,j \in G^+_k} \hat{s}_{ij}.
\end{eqnarray}

Finding the global maximum in (\ref{eq-TkTSM}) is not feasible in
general. We consider a suboptimal strategy for accomplishing
this task: sequentially construct the context by adding two genes at a
time. Start
by selecting the \emph{TSP} pair $i,j$ and setting $G_1^-=\{i\}$ and
$G_1^+=\{j\}$. Then select the pair of genes $i',j'$ distinct
from $i,j$ such that the sum of scores is maximized by
$G_2^-=\{i,i'\}$ and $G_2^+=\{j,j'\}$, that is, $\hat{\delta}_{\mathrm{TSM}}(G_k^+,G_k^-)$ is maximized over all sets $G_k^+,G_k^-$ of size
two, assuming $i \in G_k^-$ and $j \in G_k^+$. This involves
computing three new scores. Proceed in this way until $k$ pairs
have been selected.
%The one-gene version of the algorithm is similar:
%start the same way with the top-scoring pair, but then add one
%gene at a time to the pending context, alternating between the two
%sets. Here we consider the two-gene version only.

To obtain the optimal value $k^*$, one applies step (2) of the context
selection procedure, with a range of values $k \in\Omega=
\{3,5,\ldots,K\}$, for odd $K$ (the choice of $\Omega$ is dictated by
the facts that $k = 1$ reduces to 1-\textit{TSP}, whereas Proposition S1 does
not hold for even $k$):
\[
k^* = \arg\max
_{k = 3,5,\ldots,K} \hat{\tau}_{\mathrm{TSM}}\bigl(G_k^{+,*},G_k^{-,*}
\bigr),
\]
where
%
%e21 #&#
\begin{eqnarray}\label{eq-htauTSM}
&& \hat{\tau}_{\mathrm{TSM}}\bigl(G_k^{+,*},G_k^{-,*}
\bigr)\nonumber\hspace*{-10pt}
\\
&&\qquad = \frac{\hat{\delta
}_{\mathrm{TSM}}(G_k^{+,*},G_k^{-,*})}{\hat{\sigma}_{\mathrm{TSM}}(G_k^{+,*},G_k^{-,*})}\nonumber\hspace*{-10pt}
\\
&&\qquad  =
\Bigl({\widehat{E} \Bigl[\operatorname{med}\limits_{j \in
G_k^{+,*}}R_j - \operatorname{med}\limits_{i \in G_k^{-,*}}R_i
\big|  Y=1 \Bigr]
- \widehat{E} \Bigl[\operatorname{med}\limits_{j \in G_k^{+,*}}R_j -
\operatorname{med}\limits_{i \in
G_k^{-,*}}R_i   \big|  Y=0 \Bigr]}\Bigr)\hspace*{-10pt}
\\
&&\quad\qquad{} \big/
\Bigl(\widehat{\operatorname
{Var}} \Bigl(\operatorname{med}\limits_{j \in
G_k^{+,*}}R_j - \operatorname{med}\limits_{i \in G_k^{-,*}}R_i
\big|  Y=0 \Bigr)\nonumber\hspace*{-10pt}
\\
&&\hspace*{44pt}{}+\widehat{\operatorname{Var}}
\Bigl(\operatorname{med}\limits_{j \in
G_k^{+,*}}R_j - \operatorname{med}\limits_{i \in G_k^{-,*}}R_i
 \big|  Y=1 \Bigr)\Bigr)^{1/2}.\nonumber\hspace*{-10pt}
\end{eqnarray}
Notice that $\hat{\tau}_{\mathrm{TSM}}$ is defined directly by replacing
(\ref{eqTSMStatistic}) into (\ref{eqEs}) and (\ref{eq-rhos}), and
then using (\ref{eqttests}). In particular, it
does not use the approximation in (\ref{eq-TSMavg-Main}).
Finally, the optimal context is given by
$\Theta^*= (G_{k^*}^{+,*},G_{k^*}^{-,*})$.

%%% KTSP and TSM

For both \emph{KTSP} and \emph{TSM} classifiers, the step-wise process
to perform the maximization of the criterion
[cf. equations~(\ref{eq-TkKTSP})~and~(\ref{eq-TkTSM})] does not need
to be
restarted as $k$ increases, since the suboptimal contexts are nested [by
contrast, the method in \citet{kTSPPaper} employed
cross-validation to
choose $k^*$]. The detailed context selection procedure for \emph{KTSP} and \emph{TSM} classifiers is given in Algorithms S1~and~S2
in Supplement C [\citet{SuppCAOAS}].

%%%%%

% We mention that, despite the constraint of a fixed size for the
% context, the maximization in (\ref{eq-Tk}) itself may be very hard to
% accomplish exactly, due to its computational cost, in which case a
% greedy procedure is adopted whereby $\Theta_k$ is initialized with a
% small number of genes (typically the \emph{TSP}) and then genes are
% added incrementally in an iterative process until $k$ genes have been
% added. An advantage of this procedure is that, if it can be assumed
% that $\Theta^*_k \subseteq\Theta^*_l$, for $k \leq l$, i.e., the
% optimal contexts are nested, then the greedy procedure does not need
% to be restarted as $k$ increases, but can start from the previously
% obtained context. In the following sections, the details of this
% general procedure will be provided for several specific choices of
% rank discriminants.

%s3.4 #&#
\subsection{Error rates}\label{sec3.4}\label{Sec-errrt}

\begin{figure}%[t!]

\includegraphics{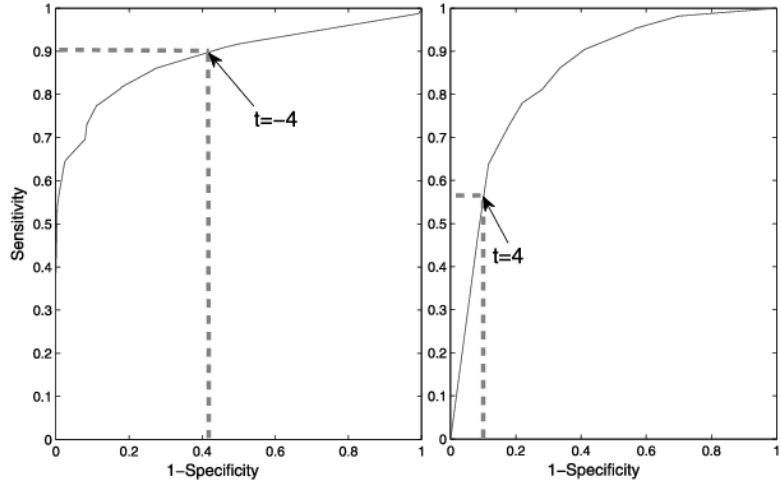}

\caption{ROC curves for \emph{TSM}.
Left: BRCA1 data. With the indicated threshold,
we can achieve sensitivity around 0.9
at the expense of specificity around 0.6.
Right: Prostate~4 data.
The given threshold reaches
0.88 specificity at the expense of sensitivity around~0.55.}\label{figROC}
\end{figure}

In this section we discuss the choice of the threshold $t$ used in
(\ref{eq-cl}). The {\it sensitivity} is defined as $P(f({\mathbf
{X}})=1 \mid
Y=1)$ and the {\it specificity} is defined as $P(f({\mathbf{X}})=0
\mid
Y=0)$. We are interested in controlling both, but trade-offs are
inevitable. The choice of which phenotype to designate as $1$ is
application-dependent; often sensitivity is relative to the more
malignant one and this is the way we have assigned labels to the
phenotypes. A given application may call for emphasizing sensitivity
at the expense of specificity or vice versa. For example, in
detecting BRCA1 mutations or with aggressive diseases such as
pancreatic cancer, high sensitivity is important, whereas for more
common and less aggressive cancers, such as prostate, it may be
preferable to limit the number of false alarms and achieve high
specificity. In principle, selecting the appropriate threshold $t$ in
(\ref{eq-cl}) allows one to achieve a desired trade-off. (A~disadvantage of \emph{TSP} is the lack of a discriminant, and thus a
procedure to adjust sensitivity and specificity.) It should be noted,
however, that in practice estimating the threshold on the training
data can be difficult; moreover, introducing a nonzero threshold
makes the decision rule somewhat more difficult to interpret. As an
example, Figure~\ref{figROC} displays the ROC curve of the \emph{TSM}
classifier for the BRCA1 and Prostate~4 studies, together with
thresholds achieving hypothetically desired scenarios.

%f2 #&#

%%% End of Section 3

%s4 #&#
\section{Experimental results}\label{sec4}\label{sectionResults}

%For each of the datasets listed in Table~\ref{TableClassSizes}, we
%applied six classification rules: \emph{TSP}, \emph{KTSP}, \emph{TSM},
%1-NN using \emph{SOS}, a linear SVM based on the $RFE$ method
%(\emph{SVM-REF}) \citet{SVMRFE2002}, and
%cross-validation.
%We employ ten repetitions of ten-fold CV and average the results, as
%recommended in \citet{BragaNeto03} and
%
%Despite the inaccuracy of small-sample cross-validation estimates (
%our interest is in obtaining a broad perspective on the relative
%comparison of the classifiers across
%many different datasets, which can be accomplished with the variant of
%CV employed here. We also discuss how
%specificity.

A summary of the rank-based discriminants developed in the preceding
sections is given in Table~\ref{TableSummary}. We learned each discriminant
for each of the data sets listed in Table~\ref{TableClassSizes}.
Among an abundance of proposed methods for high-dimensional data
classification [e.g., \citet{SVMConc}, \citet{SVMNonConv}, \citet{DWD}], we
chose two of the most effective and popular choices for predicting
phenotypes from
expression data: \emph{PAM} [\citet{PAM02}], which is a form of
\emph{LDA}, and \emph{SVM-RFE} [\citet{SVMRFE2002}], which is a form of
linear~\emph{SVM}.

Generalization errors are estimated with cross-validation, specifically
averaging the results of ten repetitions of 10-fold CV, as recommended
in \citet{BragaNeto03} and \citet{TibshiraniElementsBook}.
Despite the
inaccuracy of small-sample cross-validation estimates
[\citet{BragaNeto03}], 10-fold CV suffices to obtain the broad perspective
on relative performance across many different
data sets.

%with \emph{SOS},
% $SVM-RFE$ and $PAM$}

% To reduce computation, we filter
%the whole gene pool to the top $4000$ genes in mean rank (across both
%classes) before applying gene selection for the various rank-based
%classifiers (i.e., selecting the
%two genes for \emph{TSP}, choosing pairs for \emph{KTSP}, and the gene
%context for \emph{TSM} and \emph{SOS}).
%Notice that this does not use the class labels, and does not change
%the reported accuracies in any of the datasets.
%To be comparable in economy, we set the $SVM-RFE$ to use ten genes.
%Finally, no filtering is required for $PAM$.

%Estimated classification rates are presented in \textcolor{red}{Figures

The protocols for
training (including parameter selection) are given below. To reduce
computation, we filter the whole gene pool without using the class
labels before selecting the
context for rank discriminants (\emph{TSP}, \emph{KTSP} and \emph{TSM}). Although a variety of filtering
methods exist in the literature, such as \emph{PAM} [\citet{PAM02}],
\emph{SIS} [\citet{SIS}], Dantzig selector [\citet{Dantzig}]
and the Wilcoxon-rank
test [\citet{Wilcoxon45}], we simply use an average signal
filter: select the 4000 genes with highest mean rank
(across both classes). In particular, there is no effort to
detect ``differentially expressed'' genes. In this way we minimize the
influence of
the filtering method in assessing the performance of rank discriminants:
%No significant difference for cross-validated accuracy was seen in any
%dataset we tried without the mean rank filtering.
%Moreover, as
%mentioned before, we desire potential for mechanistical interpretation
%and mean rank can be explained as a measure to determine if a gene is
%"on" or "off" across both phenotypes. Obviously, we can roughly claim
%that mean rank filters an relatively "off" gene unable to provide a
%mechanism.
%
\begin{itemize}
\item\emph{TSP}: The single pair maximizing $s_{ij}$ over all
pairs in the $4000$ filtered genes, breaking scoring ties if necessary
with the
secondary score proposed in \citet{uTSPPaper}.

\item\emph{KTSP}: The $k$ disjoint pairs maximizing $s_{ij}$
over all pairs in the $4000$ filtered genes with the same tie-breaking
method. The number of pairs $k$ is determined
via Algorithm S1, within the range $k = 3,5,\ldots,9$, avoiding
ties in voting. Notice that $k=1$ is excluded so that \emph{KTSP} cannot
reduce to \emph{TSP}. We tried also $k = 3,5,\ldots,49$ and the
cross-validated accuracies changed insignificantly.

\item\emph{TSM}: The context is chosen from the top $4000$ genes by
the greedy selection procedure described in Algorithm S2. The size of
the two sets for computing the median rank is selected in the range $k=3,5,7,9$
(providing a unique median and thereby rendering Proposition S1
applicable). We also
tried $k = 3,5,\ldots,49$ and again the changes in the cross-validated
accuracies were insignificant. %Results are not shown for rank average
%because they are very similar, specifically one-half percent worse
%overall.

%The context is selected from top 4000 genes according to the greedy
%Algorithm 3, $3 \leq k \leq20$. (Recall that for \emph{SWP}
%$|\Theta_k|=k$ whereas $|\Theta_k|=2k$ for the others, hence the upper
%bound of $20$.) Again, $k=2$ is \emph{TSP} and we considered $k =
%3,4,\ldots,100$, but the accuracy remained approximately the same.

\item\emph{SVM-RFE}: We learned two linear \emph{SVM}s using \emph{SVM-RFE}: one with ten genes and one with a hundred genes. No
filtering was applied, since \emph{SVM-RFE} itself does that. Since
we found that the choice of the slack variable barely changes the
results, we fix $C = 0.1$. (In fact, the data are linearly separable in
nearly all loops.) Only the results for \emph{SVM-RFE}
with a hundred genes are shown since it was almost 3\% better
than with ten genes.

\item\emph{PAM}: We use the automatic filtering mechanism provided by
\citet{PAMPackage}. The prior class likelihoods were set to 0.5 and
all other parameters were set to default values. The most important
parameter is the threshold; the automatic one chosen by
the program results in relatively lower accuracy than the other methods
(84.00\%) on average. Fixing the threshold and choosing
the best one over all data sets only increases
the accuracy by one percent. Instead, for each data set and each
threshold, we estimated
the cross-validated accuracy for \emph{PAM} and report the accuracy of
the best threshold for that data set.
\end{itemize}

% % Requires \usepackage{graphicx}
% \caption{Estimated classification accuracy for predicting disease
%phenotypes. Classifiers were learned for twenty datasets and
%performance was estimated by the average of ten runs
%of ten-fold cross validation.}

%t3 #&#
\begin{table}[b]
\tabcolsep=0pt
\caption{Sensitivity/specificity for different classification
methods. Overall accuracy is calculated as the average of sensitivity and specificity}\label{tab-ss}
\begin{tabular*}{\tablewidth}{@{\extracolsep{\fill}}@{}lccccc@{}}
\hline
\textbf{Data set} & \textbf{TSP} & \textbf{TSM} & \textbf{KTSP} & \textbf{SVM} & \textbf{PAM} \\
\hline
Colon &88${}/{}$88&86${}/{}$88&87${}/{}$86&87${}/{}$73&83${}/{}$81\\
BRCA~1 &71${}/{}$75&90${}/{}$75&88${}/{}$77&68${}/{}$88&39${}/{}$82\\
CNS &41${}/{}$79&81${}/{}$88&67${}/{}$93&52${}/{}$86&77${}/{}$79\\
DLBCL &98${}/{}$97&96${}/{}$95&96${}/{}$88&97${}/{}$91&72${}/{}$100\\
Lung &92${}/{}$97&97${}/{}$99&94${}/{}$100&95${}/{}$100&97${}/{}$100\\
Marfan &82${}/{}$93&89${}/{}$90&88${}/{}$96&99${}/{}$93&88${}/{}$87\\
Crohn's &89${}/{}$90&92${}/{}$91&92${}/{}$96&100${}/{}$100&93${}/{}$98\\
Sarcoma &83${}/{}$78&88${}/{}$89&93${}/{}$91&97${}/{}$94&93${}/{}$100\\
Squamous &89${}/{}$88&88${}/{}$85&99${}/{}$92&94${}/{}$95&94${}/{}$95\\
GCM &81${}/{}$73&88${}/{}$77&90${}/{}$75&94${}/{}$80&95${}/{}$94\\
Leukemia 1 &90${}/{}$85&97${}/{}$94&97${}/{}$93&98${}/{}$97&95${}/{}$89\\
Leukemia 2 &96${}/{}$96&100${}/{}$93&100${}/{}$96&100${}/{}$96&73${}/{}$88\\
Leukemia 3 &98${}/{}$98&97${}/{}$99&97${}/{}$98&100${}/{}$100&96${}/{}$99\\
Leukemia 4 &92${}/{}$94&95${}/{}$98&96${}/{}$97&99${}/{}$97&77${}/{}$92\\
Prostate 1 &95${}/{}$93&89${}/{}$96&90${}/{}$95&91${}/{}$95&89${}/{}$91\\
Prostate 2 &68${}/{}$68&76${}/{}$79&76${}/{}$83&68${}/{}$79&77${}/{}$74\\
Prostate 3 &97${}/{}$79&99${}/{}$90&99${}/{}$83&99${}/{}$100&98${}/{}$100\\
Prostate 4 &77${}/{}$61&87${}/{}$70&86${}/{}$79&92${}/{}$62&66${}/{}$85\\
Prostate 5 &97${}/{}$99&97${}/{}$98&95${}/{}$99&100${}/{}$99&99${}/{}$100\\
Breast 1 &82${}/{}$90&82${}/{}$91&85${}/{}$91&77${}/{}$88&95${}/{}$98\\
Breast 2 &83${}/{}$82&73${}/{}$89&75${}/{}$87&71${}/{}$86&86${}/{}$88\\
% \hline
% Average & 85.59 &88.97& 88.67& 90.07& 89.92&88.19\\
\hline
\end{tabular*}
\end{table}

Table~\ref{tab-ss} shows the performance estimates of the classifiers
across 21 data sets. In addition, Figures~S1~and~S2 in Supplement B [\citet{SuppBAOAS}] display
the results in box plot format. The averages are as follows:
\emph{TSP} (85.59\%), \emph{KTSP} (90.07\%), \emph{TSM} (88.97\%), \emph{SVM-RFE} (89.92\%) and \emph{PAM} (88.19\%). The differences in the
averages among methods do not appear substantial, with the possible
exception of \emph{TSP}, which lags behind the others.

There are, however, clearly significant variations in performance within
individual data sets. In order to examine these variations at a finer
scale, possibly revealing trends to support practical recommendations,
recall that for each data set and each method, we did ten repetitions
of tenfold cross-validation, resulting in one hundred trained
classifiers and estimated rates (on the left-out subsets), which were
averaged to provide a single cross-validated classification rate. The
notch-boxes for each data set and method are plotted in Figures~S1 and S2
(Supplement B [\citet{SuppBAOAS}]). As is commonly
done, any two methods will be declared to be ``tied'' on a given data set
if the notches overlap; otherwise, that is, if the notches are disjoint,
the ``winner'' is taken to be the one with the larger median.

First, using the ``notch test'' to compare the three RIC classifiers,
\emph{KTSP} slightly outperforms \emph{TSM},
which in turn outperforms \emph{TSP}.
More specifically, \emph{KTSP} has accuracy superior to both others on ten
data sets. In terms of \emph{KTSP} vs
\emph{TSM}, \emph{KTSP} outperforms on three
data sets, vice versa on one data set and they tie on all others.
Moreover, \emph{TSM} outperforms \emph{TSP}
on nine data sets and vice versa on two
data sets. As a result, if accuracy is the dominant concern, we
recommend \emph{KTSP} among the RIC classifiers, whereas if simplicity,
transparency and links to biological mechanisms are important,
one might prefer \emph{TSP}. Comparisons with non-RIC methods (see below)
are based on \emph{KTSP}, although substituting \emph{TSM} does not
lead to appreciably different conclusions.

Second, \emph{SVM} performs better than \emph{PAM}
on six data sets and \emph{PAM} on three data sets. Hence, in the
remainder of this section we will compare \emph{KTSP} with \emph{SVM}.
We emphasize that the comparison between \emph{PAM} and \emph{SVM} is on
our particular data sets, using our particular measures of performance,
namely, cross-validation to estimate accuracy and the notch test for
pairwise comparisons. Results on other data sets or in other conditions
may differ.

Third, whereas the overall
performance statistics for \emph{KTSP} and \emph{SVM} are almost
identical, trends do emerge based on
sample size, which is obviously an important parameter and especially
useful here because it varies considerably among our data sets (Table~\ref{TableClassSizes}). To avoid fine-tuning, we only consider a
coarse and somewhat arbitrary quantization into three categories:
``small,'' ``medium'' and ``large'' data sets, defined, respectively, by
fewer than 100 (total) samples (twelve data sets), 100--200 samples
(five data sets) and more than 200 samples (four data sets). On small
data sets, \emph{KTSP} outperforms \emph{SVM} on four data sets and never
vice versa; for medium data sets, each outperforms the other on one of
the five data sets; and \emph{SVM} outperforms \emph{KTSP} on
three out of four large data sets and never vice versa.

Another criterion is sparsity: the number of
genes used by \emph{TSP} is always two and by \emph{SVM-RFE} is always
one hundred. Averaged across all data sets and loops of
cross-validation, \emph{KTSP} uses 12.5 genes, \emph{TSM} uses 10.16
genes, and \emph{PAM} uses 5771 genes.

Finally, we performed an experiment to roughly gauge the
variability in selecting the genes in the support of the various
classifiers. Taking advantage of the fact that we train 100 different
classifiers for each method and data set, each time with approximately the
same number of examples, we define a ``consistency'' measure for a pair
of classifiers as the average support overlap over all distinct pairs
of runs. That is, for any given data set and method, and any two loops
of cross-validation, let $S_1$ and $S_2$ be the supports (set of
selected genes) and define the overlap as $\frac{|S_1 \cap S_2|}{|S_1
\cup S_2|}$. This fraction is then averaged over all $100(99)/2$
pairs of loops, and obviously ranges from zero (no consistency) to one
(consistency in all loops). Whereas in 16 of the 21 data sets \emph{KTSP} had
a higher consistency score than \emph{SVM}, the more important point is that
in both cases the scores are low in absolute terms, which coheres with
other observations about the enormous variations in learned genes
signatures.
\section{Discussion and conclusions}\label{sec5}

What might be a ``mechanistic interpretation'' of the \emph{TSP}
classifier, where the
context consists of only two genes? In \citet{PricePaper}, a
reversal between the two genes Prune2 and Obscurin is shown to be an
accurate test for separating GIST and LMS. Providing an explanation,
a hypothesized mechanism, is not straightforward, although it has
been recently shown that both modulate RhoA activity (which controls
many signaling events): a splice variant of Prune2 is reported to
decrease RhoA activity when over-expressed and Obscurin contains a
Rho-GEF binding domain which helps to activate RhoA [\citet
{MRNAScenario}].

Generically, one of the most elementary regulatory motifs is simply
$A$ inhibits $B$ (denoted $A \dashv B$). For example, $A$ may be
constitutively ``on'' and $B$ constitutively ``off'' after development.
Perhaps~$A$ is a transcription factor or involved in methylation of $B$.
In the normal phenotype we see $A$ expressed, but perhaps~$A$ becomes
inactivated in the cancer phenotype, resulting in the expression of $B$,
and hence an expression reversal from normal to cancer. Still more
generally, a variety of regulatory feedback loops have been identified
in mammals. For instance, an example of a bi-stable loop is shown
below.

%f3 #&#
\begin{figure}[b]

\includegraphics{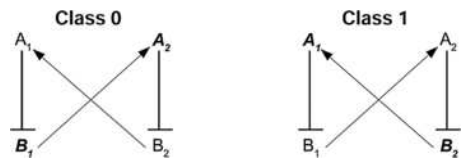}

\caption{A bi-stable feedback loop.
Molecules $A_1$, $A_2$ (resp., $B_1$, $B_2$) are from the same species, for
example, two miRNAs (resp., two mRNAs). Letters in boldface indicate an ``on'' state.}\label{figmech}
\end{figure}

Due to the activation and suppression patterns depicted in Figure~\ref{figmech}, we might expect $P(X_{A_1} < X_{A_2}|Y=0) \gg
P(X_{A_1} < X_{A_2}|Y=1)$ and $P(X_{B_1} < X_{B_2}|Y=0) \ll P(X_{B_1}
< X_{B_2}|Y=1)$. Thus, there are two expression reversals, one between
the two miRNAs and one, in the opposite direction, between the two
mRNAs. Given both miRNA and mRNA data, we might then build a
classifier based on these two switches. For example, the rank
discriminant might simply be 2-\textit{TSP}, the number of reversals
observed. Accordingly, we have argued that expression
comparisons may provide an elementary building block for a connection
between rank-based decision rules and potential mechanisms.

We have reported extensive experiments with classifiers
based on expression comparisons with different diseases and
microarray platforms and compared the results with other methods which
usually use significantly more genes. No one classifier, whether
within the rank-based collection or between them and other methods
such as \emph{SVM} and \emph{PAM}, uniformly dominates. The most appropriate
one to use is likely to be problem-dependent. Moreover, until much
larger data sets become available, it will be difficult to obtain
highly accurate estimates of generalization errors. What does seem
apparent is that our results support the conclusions reached in
earlier studies [\citet{comparclass02}, \citet{FadMicro07}, \citet{TwoGene}, \citet{Simon03}]
that simple classifiers are usually competitive with more complex ones
with microarray data and limited samples. This has important
consequences for
future developments in functional genomics since one key thrust of
``personalized medicine''
is an attempt to learn appropriate treatments for disease subtypes,
which means sample sizes will not necessarily get larger and might
even get {\it smaller}. Moreover, as attention turns increasingly
toward treatment, potentially mechanistic characterizations of
statistical decisions will become of paramount importance for
translational medicine.

%%\sname{Supplement}
%Notch-plots for Classification Accuracies,
%and Algorithms for KTSP and TSM Classifiers}
%This supplement consists of four parts: The statement and proof of
%Proposition S1; statistical tests for the assumptions made in
%Proposition S1; estimates of classification accuracies
%displayed as notch-plots for every method and dataset
%based on one hundred estimates obtained from ten runs of ten-fold
%cross-validation; and pseudo-code for the learning
%algorithms for KTSP and TSM.}
%
%proposition 1. }
%
%1.}
%
%
%and \emph{kTSP}.}
%
% \sname{Supplement A}
% \stitle{Testing Hypothesis Assumptions (A1), (A2)}
% \slink[doi]{COMPLETED BY THE TYPESETTER}
% \sdatatype{.pdf}
% \sdescription{This supplement checks the assumptions used in
%proposition 1. }
% \sname{Supplement B}
% \stitle{Proof of the proposition 1}
% \slink[doi]{COMPLETED BY THE TYPESETTER}
% \sdatatype{.pdf}
% \sdescription{This supplement provides the proof for the proposition
%1.}
% \sname{Supplement C}
% \stitle{Box-plots for the cross-validation}
% \slink[doi]{COMPLETED BY THE TYPESETTER}
% \sdatatype{.pdf}
% \sdescription{This supplement shows the box-plot figures for \emph{%TSP}, \emph{TSM}, \emph{kTSP}, \emph{SVM}, and \emph{PAM}.}
% \sname{Supplement D}
% \stitle{Pseudo-codes for RIC algorithms}
% \slink[doi]{COMPLETED BY THE TYPESETTER}
% \sdatatype{.pdf}
% \sdescription{This supplement contains the pseudo-code for \emph{TSM},
%and \emph{kTSP}.}

%This section contains the proofs of
%Propositions~1~and~2 in the text.

\begin{supplement}%[id=suppA]
\stitle{Proposition~S1}
\slink[doi]{10.1214/14-AOAS738SUPPA}
\sdatatype{.pdf}
\sfilename{AOAS738\_suppA.pdf}
\sdescription{We provide the statement and proof of Proposition S1
as well as statistical tests for the assumptions made in Proposition S1.}
\end{supplement}

\begin{supplement}%[id=suppB]
\stitle{\hspace*{-01pt}Notch-plots for classification accuracies}
\slink[doi]{10.1214/14-AOAS738SUPPB}
\sdatatype{.pdf}
\sfilename{AOAS738\_suppB.pdf}
\sdescription{We provide notch-plots of the estimates of classification
accuracy for every method and every data set based on ten runs of tenfold cross-validation.}
\end{supplement}

\begin{supplement}%[id=suppC]
\stitle{Algorithms for KTSP and TSM}
\slink[doi]{10.1214/14-AOAS738SUPPC}
\sdatatype{.pdf}
\sfilename{AOAS738\_suppC.pdf}
\sdescription{We provide a summary of the algorithms for learning the KTSP and TSM classifiers.}
\end{supplement}

% zodis "Acknowledgments" paliekamas pagal autoriu

%suskaldyti doi

% imsref loaded by linak, 2014-05-13 12:46:49
%
% imsref loaded by linak, 2014-09-12 15:13:04
% imsref loaded by linak, 2014-09-12 15:14:16
% imsref loaded by linak, 2014-09-12 15:15:59
% imsref loaded by linak, 2014-09-12 15:18:55
% imsref loaded by linak, 2014-09-12 15:21:00
% imsref loaded by linak, 2014-09-12 15:22:26
% imsref loaded by linak, 2014-09-12 15:23:08
% imsref loaded by linak, 2014-09-12 15:24:06
% imsref loaded by linak, 2014-09-12 15:30:59
% imsref loaded by linak, 2014-09-12 15:33:56

\printaddresses

\begin{thebibliography}{72}
%b1 ###
\bibitem[\protect\citeauthoryear{Afsari, Braga-Neto and
Geman}{2014a}]{SuppAAOAS}
%
\begin{bmisc}[author]
\bauthor{\bsnm{Afsari},~\bfnm{Bahman}\binits{B.}},
\bauthor{\bsnm{Braga-Neto},~\bfnm{Ulisses M.}\binits{U.~M.}} \AND
\bauthor{\bsnm{Geman},~\bfnm{Donald}\binits{D.}}
(\byear{2014}a).
\bhowpublished{Supplement to ``Rank discriminants for predicting
phenotypes from~RNA expression.'' DOI:\doiurl{10.1214/14-AOAS738SUPPA}}.
\end{bmisc}
%
\bptok{imsref}%
\endbibitem

%b2 ###
\bibitem[\protect\citeauthoryear{Afsari, Braga-Neto and
Geman}{2014b}]{SuppBAOAS}
%
\begin{bmisc}[author]
\bauthor{\bsnm{Afsari},~\bfnm{Bahman}\binits{B.}},
\bauthor{\bsnm{Braga-Neto},~\bfnm{Ulisses M.}\binits{U. M.}} \AND
\bauthor{\bsnm{Geman},~\bfnm{Donald}\binits{D.}}
(\byear{2014}b).
\bhowpublished{Supplement to ``Rank discriminants for predicting
phenotypes from~RNA expression.'' DOI:\doiurl{10.1214/14-AOAS738SUPPB}}.
\end{bmisc}
%
\bptok{imsref}%
\endbibitem

%b3 ###
\bibitem[\protect\citeauthoryear{Afsari, Braga-Neto and
Geman}{2014c}]{SuppCAOAS}
%
\begin{bmisc}[author]
\bauthor{\bsnm{Afsari},~\bfnm{Bahman}\binits{B.}},
\bauthor{\bsnm{Braga-Neto},~\bfnm{Ulisses M.}\binits{U. M.}} \AND
\bauthor{\bsnm{Geman},~\bfnm{Donald}\binits{D.}}
(\byear{2014}c).
\bhowpublished{Supplement to ``Rank discriminants for predicting
phenotypes from~RNA expression.'' DOI:\doiurl{10.1214/14-AOAS738SUPPC}}.
\end{bmisc}
%
\bptok{imsref}%
\endbibitem

%b4 ###
\bibitem[\protect\citeauthoryear{Alon et~al.}{1999}]{ColonDataSet}
\begin{barticle}[author]
\bauthor{\bsnm{Alon},~\bfnm{Ua}\binits{U.}},
\bauthor{\bsnm{Barkai},~\bfnm{Naomi}\binits{N.}},
\bauthor{\bsnm{Notterman},~\bfnm{David}\binits{D.}} \betal{et~al.}
(\byear{1999}).
\btitle{Broad patterns of gene expression revealed by clustering analysis of tumor and normal colon tissues probed by oligonucleotide arrays}.
\bjournal{Proc. Natl. Acad. Sci. USA}
\bvolume{96}
\bpages{6745--6750}.
\end{barticle}
\bptok{imsref}%
\endbibitem

%b5 ###
\bibitem[\protect\citeauthoryear{Altman et~al.}{2011}]{ClinicalLimits1}
%
\begin{barticle}[author]
\bauthor{\bsnm{Altman},~\bfnm{Ro~B.}\binits{R.~B.}},
\bauthor{\bsnm{Kroemer},~\bfnm{Ho~K.}\binits{H.~K.}},
\bauthor{\bsnm{McCarty},~\bfnm{Co~A.~}\binits{C.~A.}}
(\byear{2011}).
\btitle{Pharmacogenomics: Will the promise be fulfilled}.
\bjournal{Nat. Rev.}
\bvolume{12}
\bpages{69--73}.
\end{barticle}
%
\bptok{imsref}%
\endbibitem

%b6 ###
\bibitem[\protect\citeauthoryear{Anderson et~al.}{2007}]{kTSPProtonomics07}
\begin{barticle}[author]
\bauthor{\bsnm{Anderson},~\bfnm{Troy}\binits{T.}},
\bauthor{\bsnm{Tchernyshyov},~\bfnm{Irina}\binits{I.}},
\bauthor{\bsnm{Diez},~\bfnm{Roberto}\binits{R.}} \betal{et~al.}
(\byear{2007}).
\btitle{Discovering robust protein biomarkers for disease from relative expression reversals in 2-{D DIGE} data}.
\bjournal{Proteomics}
\bvolume{7}
\bpages{1197--1208}.
\end{barticle}
%
\bptok{imsref}%
\endbibitem

%b7 ###
\bibitem[\protect\citeauthoryear{Armstrong et~al.}{2002}]{LeukemiaG}
%
\begin{barticle}[author]
\bauthor{\bsnm{Armstrong},~\bfnm{S.~A.}\binits{S.~A.}},
\bauthor{\bsnm{Staunton},~\bfnm{J.~E.}\binits{J.~E.}},
\bauthor{\bsnm{Silverman},~\bfnm{L.~B.}\binits{L.~B.}} \betal{et~al.}
(\byear{2002}).
\btitle{MLL translocations specify a distinct gene expression profile
that distinguishes a unique leukemia}.
\bjournal{Nat. Genet.}
\bvolume{30}
\bpages{41--47}.
\end{barticle}
%
\bptok{imsref}%
\endbibitem

%b8 ###
\bibitem[\protect\citeauthoryear{Auffray}{2007}]{BioMarkers}
\begin{barticle}[author]
\bauthor{\bsnm{Auffray},~\bfnm{Ca}\binits{C.}}
(\byear{2007}).
\btitle{Protein subnetwork markers improve prediction of cancer outcome}.
\bjournal{Mol. Syst. Biol.}
\bvolume{3}
\bpages{1--2}.
\end{barticle}
%
\bptok{imsref}%
\endbibitem

%b9 ###
\bibitem[\protect\citeauthoryear{Bicciato et~al.}{2003}]{NN1}
%
\begin{barticle}[author]
\bauthor{\bsnm{Bicciato},~\bfnm{Sam}\binits{S.}},
\bauthor{\bsnm{Pandin},~\bfnm{Michael}\binits{M.}},
\bauthor{\bsnm{Didon{\`{e}}},~\bfnm{Giovanni}\binits{G.}} \AND
\bauthor{\bsnm{Bello},~\bfnm{Caro~Di}\binits{C.~D.}}
(\byear{2003}).
\btitle{Pattern identification and classification in gene expression
data using an autoassociative neural network model}.
\bjournal{Biotechnol. Bioeng.}
\bvolume{81}
\bpages{594--606}.
\end{barticle}
%
\bptok{imsref}%
\endbibitem

%b10 ###
\bibitem[\protect\citeauthoryear{Bloated, Irizarry and
Speed}{2004}]{QuantNormalization04}
%
\begin{barticle}[author]
\bauthor{\bsnm{Bloated},~\bfnm{B.}\binits{B.}},
\bauthor{\bsnm{Irizarry},~\bfnm{R.}\binits{R.}} \AND
\bauthor{\bsnm{Speed},~\bfnm{T.}\binits{T.}}
(\byear{2004}).
\btitle{A comparison of normalization methods for high density
oligonucleotide array data based on variance and bias}.
\bjournal{Bioinformatics}
\bvolume{19}
\bpages{185--193}.
\end{barticle}
%
\bptok{imsref}%
\endbibitem

%b11 ###
\bibitem[\protect\citeauthoryear{Bloom et~al.}{2004}]{NN2}
%
\begin{barticle}[author]
\bauthor{\bsnm{Bloom},~\bfnm{George}\binits{G.}},
\bauthor{\bsnm{Yang},~\bfnm{Ian}\binits{I.}},
\bauthor{\bsnm{Boulware},~\bfnm{David}\binits{D.}} \betal{et~al.}
(\byear{2004}).
\btitle{Multi-platform, multisite, microarray-based human tumor
classification.}
\bjournal{Am. J. Pathol.}
\bvolume{164}
\bpages{9--16}.
\end{barticle}
%
\bptok{imsref}%
\endbibitem

%b12 ###
\bibitem[\protect\citeauthoryear{Boulesteix, Tutz and Strimmer}{2003}]{DT1}
%
\begin{barticle}[author]
\bauthor{\bsnm{Boulesteix},~\bfnm{Abraham~L.}\binits{A.~L.}},
\bauthor{\bsnm{Tutz},~\bfnm{George.}\binits{George.}} \AND
\bauthor{\bsnm{Strimmer},~\bfnm{Kate}\binits{K.}}
(\byear{2003}).
\btitle{A CART-based approach to discover emerging patterns in
microarray data.}
\bjournal{Bioinformatics}
\bvolume{19}
\bpages{2465--2472}.
\end{barticle}
%
\bptok{imsref}%
\endbibitem

%b13 ###
\bibitem[\protect\citeauthoryear{Bradley and Mangasarian}{1998}]{SVMConc}
%
\begin{binproceedings}[author]
\bauthor{\bsnm{Bradley},~\bfnm{Paul~S.}\binits{P.~S.}} \AND
\bauthor{\bsnm{Mangasarian},~\bfnm{Olvi~L.}\binits{O.~L.}}
(\byear{1998}).
\btitle{Feature selection via voncave minimization and support vector machines}.
In \bbooktitle{ICML}
\bpages{82--90}.
\bpublisher{Morgan Kaufmann}, \blocation{Madison, WI}.
\end{binproceedings}
%
\bptok{imsref}%
\endbibitem

%b14 ###
\bibitem[\protect\citeauthoryear{Braga-Neto}{2007}]{FadMicro07}
%
\begin{barticle}[author]
\bauthor{\bsnm{Braga-Neto},~\bfnm{Ulisses~M.}\binits{U.~M.}}
(\byear{2007}).
\btitle{Fads and fallacies in the name of small-sample microarray
classification---a highlight of misunderstanding and erroneous usage in
the applications of genomic signal processing.}
\bjournal{IEEE Signal Process. Mag.}
\bvolume{24}
\bpages{91--99}.
\end{barticle}
%
\bptok{imsref}%
\endbibitem

%b15 ###
\bibitem[\protect\citeauthoryear{Braga-Neto and
Dougherty}{2004}]{BragaNeto03}
%
\begin{barticle}[pbm]
\bauthor{\bsnm{Braga-Neto},~\bfnm{Ulisses~M.}\binits{U.~M.}} \AND
\bauthor{\bsnm{Dougherty},~\bfnm{Edward~R.}\binits{E.~R.}}
(\byear{2004}).
\btitle{Is cross-validation valid for small-sample microarray classification?}
\bjournal{Bioinformatics}
\bvolume{20}
\bpages{374--380}.
\bid{doi={10.1093/bioinformatics/btg419}, issn={1367-4803},
pii={20/3/374}, pmid={14960464}}
\end{barticle}
%
\bptok{imsref}%
% NOT OUTPUTED:
% issn = 1367-4803
% number = 3
% fjournal = Bioinformatics (Oxford, England)
\endbibitem

%b16 ###
\bibitem[\protect\citeauthoryear{Buffa et~al.}{2011}]{ERGSE22220}
%
\begin{barticle}[author]
\bauthor{\bsnm{Buffa},~\bfnm{Francesca}\binits{F.}},
\bauthor{\bsnm{Camps},~\bfnm{Carme}\binits{C.}},
\bauthor{\bsnm{Winchester},~\bfnm{Laura}\binits{L.}},
\bauthor{\bsnm{Snell},~\bfnm{Cameron}\binits{C.}},
\bauthor{\bsnm{Gee},~\bfnm{Harriet}\binits{H.}},
\bauthor{\bsnm{Sheldon},~\bfnm{Helen}\binits{H.}},
\bauthor{\bsnm{Taylor},~\bfnm{Marian}\binits{M.}},
\bauthor{\bsnm{Harris},~\bfnm{Adrian}\binits{A.}} \AND
\bauthor{\bsnm{Ragoussis},~\bfnm{Jiannis}\binits{J.}}
(\byear{2011}).
\btitle{microRNA-associated progression pathways and potential
therapeutic targets identified by integrated mRNA and microRNA
expression profiling in breast cancer.}
\bjournal{Cancer Res.}
\bvolume{71}
\bpages{5635--5645}.
\end{barticle}
%
\bptok{imsref}%
\endbibitem

%b17 ###
\bibitem[\protect\citeauthoryear{Burczynski et~al.}{2006}]{Crohns}
%
\begin{barticle}[author]
\bauthor{\bsnm{Burczynski},~\bfnm{Me}\binits{M.}},
\bauthor{\bsnm{Peterson},~\bfnm{Re}\binits{R.}},
\bauthor{\bsnm{Twine},~\bfnm{Ne}\binits{N.}} \betal{et~al.}
(\byear{2006}).
\btitle{Molecular classification of Crohn's disease and ulcerative
colitis patients using transcriptional profiles in peripheral blood
mononuclear cells}.
\bjournal{Cancer Res.}
\bvolume{8}
\bpages{51--61}.
\end{barticle}
%
\bptok{imsref}%
\endbibitem

%b18 ###
\bibitem[\protect\citeauthoryear{Candes and Tao}{2007}]{Dantzig}
%
\begin{barticle}[mr]
\bauthor{\bsnm{Candes},~\bfnm{Emmanuel}\binits{E.}} \AND
\bauthor{\bsnm{Tao},~\bfnm{Terence}\binits{T.}}
(\byear{2007}).
\btitle{The {D}antzig selector: Statistical estimation when {$p$} is
much larger than {$n$}}.
\bjournal{Ann. Statist.}
\bvolume{35}
\bpages{2313--2351}.
\bid{doi={10.1214/009053606000001523}, issn={0090-5364}, mr={2382644}}
\end{barticle}
%
\bptok{imsref}%
% NOT OUTPUTED:
% issn = 0090-5364
% url = http://dx.doi.org/10.1214/009053606000001523
% number = 6
% coden = ASTSC7
% fjournal = The Annals of Statistics
\endbibitem

%b19 ###
\bibitem[\protect\citeauthoryear{Casella and Berger}{2002}]{CaseBerg02}
%
\begin{bbook}[author]
\bauthor{\bsnm{Casella},~\bfnm{G.}\binits{G.}} \AND
\bauthor{\bsnm{Berger},~\bfnm{R.~L.}\binits{R.~L.}}
(\byear{2002}).
\btitle{Statistical Inference},
\bedition{2nd} ed.
\bpublisher{Duxbury},
\blocation{Pacific Grove, CA}.
\end{bbook}
%
\bptok{imsref}%
\endbibitem

%b20 ###
\bibitem[\protect\citeauthoryear{Dettling and Buhlmann}{2003}]{Boosting1}
\begin{barticle}[author]
\bauthor{\bsnm{Dettling},~\bfnm{Michael}\binits{M.}} \AND
\bauthor{\bsnm{Buhlmann},~\bfnm{Peter}\binits{P.}}
(\byear{2003}).
\btitle{Boosting for tumor classification with gene expression data}.
\bjournal{Bioinformatics}
\bvolume{19}
\bpages{1061--1069}.
\end{barticle}
%
\bptok{imsref}%
\endbibitem

%b21 ###
\bibitem[\protect\citeauthoryear{Dudoit, Fridlyand and
Speed}{2002}]{comparclass02}
%
\begin{barticle}[mr]
\bauthor{\bsnm{Dudoit},~\bfnm{Sandrine}\binits{S.}},
\bauthor{\bsnm{Fridlyand},~\bfnm{Jane}\binits{J.}} \AND
\bauthor{\bsnm{Speed},~\bfnm{Terence~P.}\binits{T.~P.}}
(\byear{2002}).
\btitle{Comparison of discrimination methods for the classification of
tumors using gene expression data}.
\bjournal{J. Amer. Statist. Assoc.}
\bvolume{97}
\bpages{77--87}.
\bid{doi={10.1198/016214502753479248}, issn={0162-1459}, mr={1963389}}
\end{barticle}
%
\bptok{imsref}%
% NOT OUTPUTED:
% issn = 0162-1459
% url = http://dx.doi.org/10.1198/016214502753479248
% number = 457
% coden = JSTNAL
% fjournal = Journal of the American Statistical Association
\endbibitem

%b22 ###
\bibitem[\protect\citeauthoryear{Edelman et~al.}{2009}]{Edelman09}
%
\begin{barticle}[author]
\bauthor{\bsnm{Edelman},~\bfnm{Lucas}\binits{L.}},
\bauthor{\bsnm{Toia},~\bfnm{Giuseppe}\binits{G.}},
\bauthor{\bsnm{Geman},~\bfnm{Donald}\binits{D.}} \betal{et~al.}
(\byear{2009}).
\btitle{Two-transcript gene expression classifiers in the diagnosis
and prognosis of human diseases}.
\bjournal{BMC Genomics}
\bvolume{10}
\bpages{583}.
\end{barticle}
%
\bptok{imsref}%
\endbibitem

%b23 ###
\bibitem[\protect\citeauthoryear{Enerly et~al.}{2011}]{ERGSE19783}
%
\begin{barticle}[author]
\bauthor{\bsnm{Enerly},~\bfnm{Espen}\binits{E.}},
\bauthor{\bsnm{Steinfeld},~\bfnm{Israel}\binits{I.}},
\bauthor{\bsnm{Kleivi},~\bfnm{Kristine}\binits{K.}},
\bauthor{\bsnm{Leivonen},~\bfnm{Suvi-Katri}\binits{S.-K.}} \betal{et~al.}
(\byear{2011}).
\btitle{miRNA--mRNA integrated analysis reveals roles for miRNAs in
primary breast tumors}.
\bjournal{PLoS ONE}
\bvolume{6}
\bpages{0016915}.
\end{barticle}
%
\bptok{imsref}%
\endbibitem

%b24 ###
\bibitem[\protect\citeauthoryear{Evans et~al.}{2011}]{ClinicalLimits3}
%
\begin{barticle}[author]
\bauthor{\bsnm{Evans},~\bfnm{James~P.}\binits{J.~P.}},
\bauthor{\bsnm{Meslin},~\bfnm{Ed~M.}\binits{E.~M.}},
\bauthor{\bsnm{Marteau},~\bfnm{Te~M.}\binits{T.~M.}} \AND
\bauthor{\bsnm{Caulfield},~\bfnm{Te}\binits{T.}}
(\byear{2011}).
\btitle{Deflating the genomic bubble}.
\bjournal{Science}
\bvolume{331}
\bpages{861--862}.
\end{barticle}
%
\bptok{imsref}%
\endbibitem

%b25 ###
\bibitem[\protect\citeauthoryear{Fan and Fan}{2008}]{fan2008high}
%
\begin{barticle}[mr]
\bauthor{\bsnm{Fan},~\bfnm{Jianqing}\binits{J.}} \AND
\bauthor{\bsnm{Fan},~\bfnm{Yingying}\binits{Y.}}
(\byear{2008}).
\btitle{High-dimensional classification using features annealed
independence rules}.
\bjournal{Ann. Statist.}
\bvolume{36}
\bpages{2605--2637}.
\bid{doi={10.1214/07-AOS504}, issn={0090-5364}, mr={2485009}}
\end{barticle}
%
\bptok{imsref}%
% NOT OUTPUTED:
% issn = 0090-5364
% url = http://dx.doi.org/10.1214/07-AOS504
% number = 6
% coden = ASTSC7
% fjournal = The Annals of Statistics
\endbibitem

%b26 ###
\bibitem[\protect\citeauthoryear{Fan and Lv}{2008}]{SIS}
%
\begin{barticle}[mr]
\bauthor{\bsnm{Fan},~\bfnm{Jianqing}\binits{J.}} \AND
\bauthor{\bsnm{Lv},~\bfnm{Jinchi}\binits{J.}}
(\byear{2008}).
\btitle{Sure independence screening for ultrahigh dimensional feature space}.
\bjournal{J.~R. Stat. Soc. Ser. B Stat. Methodol.}
\bvolume{70}
\bpages{849--911}.
\bid{doi={10.1111/j.1467-9868.2008.00674.x}, issn={1369-7412}, mr={2530322}}
\end{barticle}
%
\bptok{imsref}%
% NOT OUTPUTED:
% issn = 1369-7412
% url = http://dx.doi.org/10.1111/j.1467-9868.2008.00674.x
% number = 5
% fjournal = Journal of the Royal Statistical Society. Series B.
%Statistical Methodology
\endbibitem

%b27 ###
\bibitem[\protect\citeauthoryear{Funk}{2012}]{MRNAScenario}
%
\begin{bmisc}[author]
\bauthor{\bsnm{Funk},~\bfnm{Cory}\binits{C.}}
(\byear{2012}).
\bhowpublished{Personal communication. Institute for Systems Biology,
Seattle, WA}.
\end{bmisc}
%
\bptok{imsref}%
% NOT OUTPUTED:
% sortkey = Funk(2012
\endbibitem

%b28 ###
\bibitem[\protect\citeauthoryear{Geman et~al.}{2004}]{Geman04}
\begin{barticle}[mr]
\bauthor{\bsnm{Geman},~\bfnm{Donald}\binits{D.}},
\bauthor{\bsnm{d'Avignon},~\bfnm{Christian}\binits{C.}},
\bauthor{\bsnm{Naiman},~\bfnm{Daniel~Q.}\binits{D.~Q.}} \AND
\bauthor{\bsnm{Winslow},~\bfnm{Raimond~L.}\binits{R.~L.}}
(\byear{2004}).
\btitle{Classifying gene expression profiles from pairwise m{RNA} comparisons}.
\bjournal{Stat. Appl. Genet. Mol. Biol.}
\bvolume{3}
\bpages{Art. 19, 21~pp. (electronic)}.
\bid{doi={10.2202/1544-6115.1071}, issn={1544-6115}, mr={2101468}}
\end{barticle}
%
\bptok{imsref}%
\endbibitem

%b29 ###
\bibitem[\protect\citeauthoryear{Geman et~al.}{2008}]{ICMLA08}
%
\begin{binproceedings}[author]
\bauthor{\bsnm{Geman},~\bfnm{Donald}\binits{D.}},
\bauthor{\bsnm{Afsari},~\bfnm{Bahman}\binits{B.}},
\bauthor{\bsnm{Tan},~\bfnm{Aik~Choon}\binits{A.~C.}} \AND
\bauthor{\bsnm{Naiman},~\bfnm{Daniel~Q.}\binits{D.~Q.}}
(\byear{2008}).
\btitle{Microarray classification from several two-gene expression
comparisons}.
In \bbooktitle{Machine Learning and Applications, 2008. ICMLA'08.
Seventh International Conference}
\bpages{583--585}.
\bpublisher{IEEE}, \blocation{San Diego, CA}.
%Competition)}.
\end{binproceedings}
%
\bptok{imsref}%
\endbibitem

%b30 ###
\bibitem[\protect\citeauthoryear{Golub et~al.}{1999}]{LeukemiaDataSet}
%
\begin{barticle}[author]
\bauthor{\bsnm{Golub},~\bfnm{Tom~R.}\binits{T.~R.}},
\bauthor{\bsnm{Slonim},~\bfnm{David~K.}\binits{D.~K.}},
\bauthor{\bsnm{Tamayo},~\bfnm{Peter}\binits{P.}} \betal{et~al.}
(\byear{1999}).
\btitle{Molecular classification of cancer: Class discovery and class
prediction by gene expression monitoring}.
\bjournal{Science}
\bvolume{286}
\bpages{531--537}.
\end{barticle}
%
\bptok{imsref}%
\endbibitem

%b31 ###
\bibitem[\protect\citeauthoryear{Gordon et~al.}{2002}]{LungCancer}
%
\begin{barticle}[pbm]
\bauthor{\bsnm{Gordon},~\bfnm{Gavin~J.}\binits{G.~J.}},
\bauthor{\bsnm{Jensen},~\bfnm{Roderick~V.}\binits{R.~V.}},
\bauthor{\bsnm{Hsiao},~\bfnm{Li-Li}\binits{L.-L.}},
\bauthor{\bsnm{Gullans},~\bfnm{Steven~R.}\binits{S.~R.}},
\bauthor{\bsnm{Blumenstock},~\bfnm{Joshua~E.}\binits{J.~E.}},
\bauthor{\bsnm{Ramaswamy},~\bfnm{Sridhar}\binits{S.}},
\bauthor{\bsnm{Richards},~\bfnm{William~G.}\binits{W.~G.}},
\bauthor{\bsnm{Sugarbaker},~\bfnm{David~J.}\binits{D.~J.}} \AND
\bauthor{\bsnm{Bueno},~\bfnm{Raphael}\binits{R.}}
(\byear{2002}).
\btitle{Translation of microarray data into clinically relevant cancer
diagnostic tests using gene expression ratios in lung cancer and mesothelioma}.
\bjournal{Cancer Res.}
\bvolume{62}
\bpages{4963--4967}.
\bid{issn={0008-5472}, pmid={12208747}}
\end{barticle}
%
\bptok{imsref}%
% NOT OUTPUTED:
% issn = 0008-5472
% number = 17
% fjournal = Cancer research
\endbibitem

%b32 ###
\bibitem[\protect\citeauthoryear{Guo, Hastie and
Tibshirani}{2005}]{guo2005regularized}
%
\begin{barticle}[author]
\bauthor{\bsnm{Guo},~\bfnm{Yaqian}\binits{Y.}},
\bauthor{\bsnm{Hastie},~\bfnm{Trevor}\binits{T.}} \AND
\bauthor{\bsnm{Tibshirani},~\bfnm{Robert}\binits{R.}}
(\byear{2005}).
\btitle{Regularized discriminant analysis and its application in microarrays}.
\bjournal{Biostatistics}
\bvolume{1}
\bpages{1--18}.
\end{barticle}
%
\bptok{imsref}%
\endbibitem

%b33 ###
\bibitem[\protect\citeauthoryear{Guo, Hastie and Tibshirani}{2007}]{SCRDA07}
%
\begin{barticle}[pbm]
\bauthor{\bsnm{Guo},~\bfnm{Yaqian}\binits{Y.}},
\bauthor{\bsnm{Hastie},~\bfnm{Trevor}\binits{T.}} \AND
\bauthor{\bsnm{Tibshirani},~\bfnm{Robert}\binits{R.}}
(\byear{2007}).
\btitle{Regularized linear discriminant analysis and its application
in microarrays}.
\bjournal{Biostatistics}
\bvolume{8}
\bpages{86--100}.
\bid{doi={10.1093/biostatistics/kxj035}, issn={1465-4644},
pii={kxj035}, pmid={16603682}}
\end{barticle}
%
\bptok{imsref}%
% NOT OUTPUTED:
% issn = 1465-4644
% number = 1
% fjournal = Biostatistics (Oxford, England)
\endbibitem

%b34 ###
\bibitem[\protect\citeauthoryear{Guyon et~al.}{2002}]{SVMRFE2002}
%
\begin{barticle}[author]
\bauthor{\bsnm{Guyon},~\bfnm{Isabelle}\binits{I.}},
\bauthor{\bsnm{Weston},~\bfnm{Jason}\binits{J.}},
\bauthor{\bsnm{Barnhill},~\bfnm{Stephen}\binits{S.}} \AND
\bauthor{\bsnm{Vapnik},~\bfnm{Vladimir}\binits{V.}}
(\byear{2002}).
\btitle{Gene selection for cancer classification using support vector
machines}.
\bjournal{Mach. Learn.}
\bvolume{46}
\bpages{389--422}.
\end{barticle}
%
\bptok{imsref}%
\endbibitem

%b35 ###
\bibitem[\protect\citeauthoryear{Hastie, Tibshirani and
Friedman}{2001}]{TibshiraniElementsBook}
%
\begin{bbook}[mr]
\bauthor{\bsnm{Hastie},~\bfnm{Trevor}\binits{T.}},
\bauthor{\bsnm{Tibshirani},~\bfnm{Robert}\binits{R.}} \AND
\bauthor{\bsnm{Friedman},~\bfnm{Jerome}\binits{J.}}
(\byear{2001}).
\btitle{The Elements of Statistical Learning: Data Mining, Inference,
and Prediction}.
\bseries{Springer Series in Statistics}.
\bpublisher{Springer},
\blocation{New York}.
\bid{doi={10.1007/978-0-387-21606-5}, mr={1851606}}
\end{bbook}
%
\bptok{imsref}%
% NOT OUTPUTED:
% isbn = 0-387-95284-5
% url = http://dx.doi.org/10.1007/978-0-387-21606-5
% fpage = xvi+533
\endbibitem

%b36 ###
\bibitem[\protect\citeauthoryear{Jones et~al.}{2008}]{Vogelestein08}
%
\begin{barticle}[author]
\bauthor{\bsnm{Jones},~\bfnm{Si{\^{a}}n}\binits{S.}},
\bauthor{\bsnm{Zhang},~\bfnm{Xiaosong}\binits{X.}},
\bauthor{\bsnm{Parsons},~\bfnm{D.~Williams}\binits{D.~W.}} \betal{et~al.}
(\byear{2008}).
\btitle{Core signaling pathways in human pancreatic cancers revealed
by global genomic analyses}.
\bjournal{Science}
\bvolume{321}
\bpages{1801--1806}.
\end{barticle}
%
\bptok{imsref}%
\endbibitem

%b37 ###
\bibitem[\protect\citeauthoryear{Khan et~al.}{2001}]{NN3}
\begin{barticle}[author]
\bauthor{\bsnm{Khan},~\bfnm{Javad}\binits{J.}},
\bauthor{\bsnm{Wei},~\bfnm{Jason~S.}\binits{J.~S.}},
\bauthor{\bsnm{Ringn{\'{e}}r},~\bfnm{Michael}\binits{M.}} \betal{et~al.}
(\byear{2001}).
\btitle{Classification and diagnostic prediction of cancers using gene expression profiling and artificial neural networks.}
\bjournal{Nat. Med.}
\bvolume{7}
\bpages{673--679}.
\end{barticle}
%
%%
%(\byear{2001}).
%expression profiling and artificial neural networks.}
%%
\bptok{imsref}%
\endbibitem

%b38 ###
\bibitem[\protect\citeauthoryear{Kohlmann et~al.}{2008}]{Leukemia3}
%
\begin{barticle}[author]
\bauthor{\bsnm{Kohlmann},~\bfnm{Alexander}\binits{A.}},
\bauthor{\bsnm{Kipps},~\bfnm{Thomas~J.}\binits{T.~J.}},
\bauthor{\bsnm{Rassenti},~\bfnm{Laura~Z.}\binits{L.~Z.}} \AND
\bauthor{\bsnm{Downing},~\bfnm{James~R.}\binits{J.~R.}}
(\byear{2008}).
\btitle{An international standardization programme towards the
application of
gene expression profiling in routine leukaemia diagnostics: The
microarray innovations in leukemia study prephase}.
\bjournal{Br. J. Haematol.}
\bvolume{142}
\bpages{802--807}.
\end{barticle}
%
\bptok{imsref}%
\endbibitem

%b39 ###
\bibitem[\protect\citeauthoryear{Kuriakose, Chen
et~al.}{2004}]{SquamousDataSet}
%
\begin{barticle}[author]
\bauthor{\bsnm{Kuriakose},~\bfnm{Mo~A.}\binits{M.~A.}},
\bauthor{\bsnm{Chen},~\bfnm{Wo~T.}\binits{W.~T.}} \betal{et~al.}
(\byear{2004}).
\btitle{Selection and validation of differentially expressed genes in
head and neck cancer}.
\bjournal{Cell. Mol. Life Sci.}
\bvolume{61}
\bpages{1372--1383}.
\end{barticle}
%
\bptok{imsref}%
\endbibitem

%b40 ###
\bibitem[\protect\citeauthoryear{Lee et~al.}{2008}]{Network1}
%
\begin{barticle}[author]
\bauthor{\bsnm{Lee},~\bfnm{Ed}\binits{E.}},
\bauthor{\bsnm{Chuang},~\bfnm{H.~Y.}\binits{H.~Y.}},
\bauthor{\bsnm{Kim},~\bfnm{J.~W.}\binits{J.~W.}} \betal{et~al.}
(\byear{2008}).
\btitle{Inferring pathway activity toward precise disease classification.}
\bjournal{PLoS Comput. Biol.}
\bvolume{4}
\bpages{e1000217}.
\end{barticle}
%
\bptok{imsref}%
\endbibitem

%b41 ###
\bibitem[\protect\citeauthoryear{Lin et~al.}{2009}]{TSTPaper}
%
\begin{barticle}[author]
\bauthor{\bsnm{Lin},~\bfnm{Xue}\binits{X.}},
\bauthor{\bsnm{Afsari},~\bfnm{Bahman}\binits{B.}},
\bauthor{\bsnm{Marchionni},~\bfnm{Luigi}\binits{L.}} \betal{et~al.}
(\byear{2009}).
\btitle{The ordering of expression among a few genes can provide
simple cancer biomarkers and signal BRCA1 mutations}.
\bjournal{BMC Bioinformatics}
\bvolume{10}
\bpages{256}.
\end{barticle}
%
\bptok{imsref}%
\endbibitem

%b42 ###
\bibitem[\protect\citeauthoryear{Ma et~al.}{2004}]{Ma04}
%
\begin{barticle}[author]
\bauthor{\bsnm{Ma},~\bfnm{Xiao~Jun}\binits{X.~J.}},
\bauthor{\bsnm{Wang},~\bfnm{Zuncai}\binits{Z.}},
\bauthor{\bsnm{Ryan},~\bfnm{Paula~D.}\binits{P.~D.}} \betal{et~al.}
(\byear{2004}).
\btitle{A two-gene expression ratio predicts clinical outcome in
breast cancer patients treated with tamoxifen}.
\bjournal{Cancer Cell}
\bvolume{5}
\bpages{607--616}.
\end{barticle}
%
\bptok{imsref}%
\endbibitem

%b43 ###
\bibitem[\protect\citeauthoryear{Marron, Todd and Ahn}{2007}]{DWD}
%
\begin{barticle}[mr]
\bauthor{\bsnm{Marron},~\bfnm{J.~S.}\binits{J.~S.}},
\bauthor{\bsnm{Todd},~\bfnm{Michael~J.}\binits{M.~J.}} \AND
\bauthor{\bsnm{Ahn},~\bfnm{Jeongyoun}\binits{J.}}
(\byear{2007}).
\btitle{Distance-weighted discrimination}.
\bjournal{J. Amer. Statist. Assoc.}
\bvolume{102}
\bpages{1267--1271}.
\bid{doi={10.1198/016214507000001120}, issn={0162-1459}, mr={2412548}}
\end{barticle}
%
\bptok{imsref}%
% NOT OUTPUTED:
% issn = 0162-1459
% url = http://dx.doi.org/10.1198/016214507000001120
% number = 480
% coden = JSTNAL
% fjournal = Journal of the American Statistical Association
\endbibitem

%b44 ###
\bibitem[\protect\citeauthoryear{Marshall}{2011}]{ClinicalLimits2}
%
\begin{barticle}[author]
\bauthor{\bsnm{Marshall},~\bfnm{Ed}\binits{E.}}
(\byear{2011}).
\btitle{Waiting for the revolution}.
\bjournal{Science}
\bvolume{331}
\bpages{526--529}.
\end{barticle}
%
\bptok{imsref}%
\endbibitem

%b45 ###
\bibitem[\protect\citeauthoryear{Mills et~al.}{2009}]{Leukemia4}
\begin{barticle}[author]
\bauthor{\bsnm{Mills},~\bfnm{K.~I.}\binits{K.~I.}},
\bauthor{\bsnm{Kohlmann},~\bfnm{A.}\binits{A.}},
\bauthor{\bsnm{Williams},~\bfnm{P.~M.}\binits{P.~M.}},
\bauthor{\bsnm{Wieczorek},~\bfnm{L.}\binits{L.}} \betal{et~al.}
(\byear{2009}).
\btitle{Microarray-based classifiers and prognosis models identify subgroups with distinct clinical outcomes and high risk of AML transformation of myelodysplastic syndrome}.
\bjournal{Blood}
\bvolume{114}
\bpages{1063--1072}.
\end{barticle}
%
%%
%(\byear{2009}).
%with distinct clinical outcomes and high risk of AML transformation of
%myelodysplastic syndrome}.
%%
\bptok{imsref}%
\endbibitem

%b46 ###
\bibitem[\protect\citeauthoryear{Peng et~al.}{2003}]{SVM1}
%
\begin{barticle}[author]
\bauthor{\bsnm{Peng},~\bfnm{Se}\binits{S.}},
\bauthor{\bsnm{Xu},~\bfnm{Qe}\binits{Q.}},
\bauthor{\bsnm{Ling},~\bfnm{Xe}\binits{X.}} \betal{et~al.}
(\byear{2003}).
\btitle{Molecular classification of cancer types from microarray data
using the combination of genetic algorithms and support vector machines}.
\bjournal{FEBS Lett.}
\bvolume{555}
\bpages{358--362}.
\end{barticle}
%
\bptok{imsref}%
\endbibitem

%b47 ###
\bibitem[\protect\citeauthoryear{Pomeroy et~al.}{2002}]{CNSDataSet}
%
\begin{barticle}[author]
\bauthor{\bsnm{Pomeroy},~\bfnm{Cott}\binits{C.}},
\bauthor{\bsnm{Tamayo},~\bfnm{Pablo}\binits{P.}},
\bauthor{\bsnm{Gaasenbeek},~\bfnm{Michelle}\binits{M.}} \betal{et~al.}
(\byear{2002}).
\btitle{Prediction of central nervous system embryonal tumour outcome
based on gene expression}.
\bjournal{Nature}
\bvolume{415}
\bpages{436--442}.
\end{barticle}
%
\bptok{imsref}%
\endbibitem

%b48 ###
\bibitem[\protect\citeauthoryear{Price et~al.}{2007}]{PricePaper}
\begin{barticle}[author]
\bauthor{\bsnm{Price},~\bfnm{Nathan}\binits{N.}},
\bauthor{\bsnm{Trent},~\bfnm{Jonathan}\binits{J.}},
\bauthor{\bsnm{El-Naggar},~\bfnm{Adel}\binits{A.}} \betal{et~al.}
(\byear{2007}).
\btitle{Highly accurate two-gene classifier for differentiating gastrointestinal stromal tumors and leimyosarcomas.}
\bjournal{Proc. Natl. Acad. Sci. USA}
\bvolume{43}
\bpages{3414--3419}.
\end{barticle}
%
%%
%(\byear{2007}).
%gastrointestinal stromal tumors and leimyosarcomas.}
%%
\bptok{imsref}%
\endbibitem

%b49 ###
\bibitem[\protect\citeauthoryear{Qu et~al.}{2002}]{Boosting2}
%
\begin{barticle}[author]
\bauthor{\bsnm{Qu},~\bfnm{Yo}\binits{Y.}},
\bauthor{\bsnm{Adam},~\bfnm{Bo}\binits{B.}},
\bauthor{\bsnm{Yasui},~\bfnm{Yo}\binits{Y.}} \betal{et~al.}
(\byear{2002}).
\btitle{Boosted decision tree analysis of surface-enhanced laser
desorption/ionization mass spectral serum profiles discriminates
prostate cancer from noncancer patients}.
\bjournal{Clin. Chem.}
\bvolume{48}
\bpages{1835--1843}.
\end{barticle}
%
\bptok{imsref}%
\endbibitem

%b50 ###
\bibitem[\protect\citeauthoryear{Ramaswamy et~al.}{2001}]{GCMDataSet}
%
\begin{barticle}[author]
\bauthor{\bsnm{Ramaswamy},~\bfnm{Se}\binits{S.}},
\bauthor{\bsnm{Tamayo},~\bfnm{Pe}\binits{P.}},
\bauthor{\bsnm{Rifkin},~\bfnm{Re}\binits{R.}} \betal{et~al.}
(\byear{2001}).
\btitle{Multiclass cancer diagnosis using tumor gene expression signatures}.
\bjournal{Proc. Natl. Acad. Sci. USA}
\bvolume{98}
\bpages{15149--15154}.
\end{barticle}
%
\bptok{imsref}%
\endbibitem

%b51 ###
\bibitem[\protect\citeauthoryear{Raponi et~al.}{2008}]{BloodTSP}
%
\begin{barticle}[author]
\bauthor{\bsnm{Raponi},~\bfnm{Mitch}\binits{M.}},
\bauthor{\bsnm{Lancet},~\bfnm{Jeffrey~E.}\binits{J.~E.}},
\bauthor{\bsnm{Fan},~\bfnm{Hongtao}\binits{H.}} \betal{et~al.}
(\byear{2008}).
\btitle{A 2-gene classifier for predicting response to the
farnesyltransferase inhibitor tipifarnib in acute myeloid leukemia}.
\bjournal{Blood}
\bvolume{111}
\bpages{2589--2596}.
\end{barticle}
%
\bptok{imsref}%
\endbibitem

%b52 ###
\bibitem[\protect\citeauthoryear{Shipp et~al.}{2002}]{DLBCLDataSet}
%
\begin{barticle}[author]
\bauthor{\bsnm{Shipp},~\bfnm{Ma}\binits{M.}},
\bauthor{\bsnm{Ross},~\bfnm{Kn}\binits{K.}},
\bauthor{\bsnm{Tamayo},~\bfnm{Po}\binits{P.}} \betal{et~al.}
(\byear{2002}).
\btitle{Diffuse large B-cell lymphoma outcome prediction by
gene-expression profiling and supervised machine learning}.
\bjournal{Nat. Med.}
\bvolume{8}
\bpages{68--74}.
\end{barticle}
%
\bptok{imsref}%
\endbibitem

%b53 ###
\bibitem[\protect\citeauthoryear{Simon et~al.}{2003}]{Simon03}
%
\begin{barticle}[pbm]
\bauthor{\bsnm{Simon},~\bfnm{Richard}\binits{R.}},
\bauthor{\bsnm{Radmacher},~\bfnm{Michael~D.}\binits{M.~D.}},
\bauthor{\bsnm{Dobbin},~\bfnm{Kevin}\binits{K.}} \AND
\bauthor{\bsnm{McShane},~\bfnm{Lisa~M.}\binits{L.~M.}}
(\byear{2003}).
\btitle{Pitfalls in the use of DNA microarray data for diagnostic and
prognostic classification}.
\bjournal{J. Natl. Cancer Inst.}
\bvolume{95}
\bpages{14--18}.
\bid{issn={0027-8874}, pmid={12509396}}
\end{barticle}
%
\bptok{imsref}%
% NOT OUTPUTED:
% issn = 0027-8874
% number = 1
% fjournal = Journal of the National Cancer Institute
\endbibitem

%b54 ###
\bibitem[\protect\citeauthoryear{Singh et~al.}{2002}]{Prostate1}
%
\begin{barticle}[author]
\bauthor{\bsnm{Singh},~\bfnm{Dheeraj}\binits{D.}},
\bauthor{\bsnm{Febbo},~\bfnm{Pablo}\binits{P.}},
\bauthor{\bsnm{Ross},~\bfnm{Keith}\binits{K.}} \betal{et~al.}
(\byear{2002}).
\btitle{Gene expression correlates of clinical prostate cancer behavior}.
\bjournal{Cancer Cell}
\bvolume{1}
\bpages{203--209}.
\end{barticle}
%
\bptok{imsref}%
\endbibitem

%b55 ###
\bibitem[\protect\citeauthoryear{Stuart et~al.}{2004}]{Prostate2}
%
\begin{barticle}[author]
\bauthor{\bsnm{Stuart},~\bfnm{Ro}\binits{R.}},
\bauthor{\bsnm{Wachsman},~\bfnm{Wo}\binits{W.}},
\bauthor{\bsnm{Berry},~\bfnm{Co}\binits{C.}} \betal{et~al.}
(\byear{2004}).
\btitle{In silico dissection of cell-type-associated patterns of gene
expression in prostate cancer}.
\bjournal{Proc. Natl. Acad. Sci. USA}
\bvolume{101}
\bpages{615--620}.
\end{barticle}
%
\bptok{imsref}%
\endbibitem

%b56 ###
\bibitem[\protect\citeauthoryear{Tan et~al.}{2005}]{kTSPPaper}
%
\begin{barticle}[author]
\bauthor{\bsnm{Tan},~\bfnm{Aik~Choon}\binits{A.~C.}},
\bauthor{\bsnm{Naiman},~\bfnm{Daniel~Q.}\binits{D.~Q.}},
\bauthor{\bsnm{Xu},~\bfnm{Lei}\binits{L.}} \betal{et~al.}
(\byear{2005}).
\btitle{Simple decision rules for classifying human cancers from gene
expression profiles}.
\bjournal{Bioinformatics}
\bvolume{21}
\bpages{3896--3904}.
\end{barticle}
%
\bptok{imsref}%
\endbibitem

%b57 ###
\bibitem[\protect\citeauthoryear{Thomas et~al.}{2007}]{MutationOncogenes}
%
\begin{barticle}[author]
\bauthor{\bsnm{Thomas},~\bfnm{Roman~K.}\binits{R.~K.}},
\bauthor{\bsnm{Baker},~\bfnm{Alissa~C.}\binits{A.~C.}},
\bauthor{\bsnm{DeBiasi},~\bfnm{Ralph~M.}\binits{R.~M.}} \betal{et~al.}
(\byear{2007}).
\btitle{High-throughput oncogene mutation profiling in human cancer}.
\bjournal{Nature Genetics}
\bvolume{39}
\bpages{347--351}.
\end{barticle}
%
\bptok{imsref}%
\endbibitem

%b58 ###
\bibitem[\protect\citeauthoryear{Tibshirani}{2011}]{PAMPackage}
%
\begin{bmisc}[author]
\bauthor{\bsnm{Tibshirani},~\bfnm{Robert}\binits{R.}}
(\byear{2011}).
\bhowpublished{PAM R Package. Available at \url
{http://www-stat.stanford.edu/\textasciitilde tibs/PAM/Rdist/index.html}.}
\end{bmisc}
%
\bptok{imsref}%
% NOT OUTPUTED:
% sortkey = Tibshirani(2011
\endbibitem

%b59 ###
\bibitem[\protect\citeauthoryear{Tibshirani et~al.}{2002}]{PAM02}
%
\begin{barticle}[author]
\bauthor{\bsnm{Tibshirani},~\bfnm{Robert}\binits{R.}},
\bauthor{\bsnm{Hastie},~\bfnm{Trevor}\binits{T.}},
\bauthor{\bsnm{Narasimhan},~\bfnm{Balasubramanian}\binits{B.}} \AND
\bauthor{\bsnm{Chu},~\bfnm{Gilbert}\binits{G.}}
(\byear{2002}).
\btitle{Diagnosis of multiple cancer types by shrunken centroids of
gene expression}.
\bjournal{Proc. Natl. Acad. Sci. USA}
\bvolume{99}
\bpages{6567--6572}.
\end{barticle}
%
\bptok{imsref}%
\endbibitem

%b60 ###
\bibitem[\protect\citeauthoryear{Wang}{2012}]{TwoGene}
%
\begin{barticle}[pbm]
\bauthor{\bsnm{Wang},~\bfnm{Xiaosheng}\binits{X.}}
(\byear{2012}).
\btitle{Robust two-gene classifiers for cancer prediction}.
\bjournal{Genomics}
\bvolume{99}
\bpages{90--95}.
\bid{doi={10.1016/j.ygeno.2011.11.003}, issn={1089-8646},
mid={NIHMS340704}, pii={S0888-7543(11)00260-6}, pmcid={3273650},
pmid={22138042}}
\end{barticle}
%
\bptok{imsref}%
% NOT OUTPUTED:
% issn = 1089-8646
% number = 2
\endbibitem

%b61 ###
\bibitem[\protect\citeauthoryear{Weichselbaum
et~al.}{2008}]{BreastWeichselbaum}
%
\begin{barticle}[author]
\bauthor{\bsnm{Weichselbaum},~\bfnm{Ralph~R.}\binits{R.~R.}},
\bauthor{\bsnm{Ishwaranc},~\bfnm{Hemant}\binits{H.}},
\bauthor{\bsnm{Yoona},~\bfnm{Taewon}\binits{T.}} \betal{et~al.}
(\byear{2008}).
\btitle{An interferon-related gene signature for DNA damage resistance
is a predictive marker for chemotherapy and radiation for breast cancer}.
\bjournal{Proc. Natl. Acad. Sci. USA}
\bvolume{105}
\bpages{18490--18495}.
\end{barticle}
%
\bptok{imsref}%
\endbibitem

%b62 ###
\bibitem[\protect\citeauthoryear{Welsh et~al.}{2001}]{Prostate3}
%
\begin{barticle}[author]
\bauthor{\bsnm{Welsh},~\bfnm{Jo}\binits{J.}},
\bauthor{\bsnm{Sapinoso},~\bfnm{Lo}\binits{L.}},
\bauthor{\bsnm{Su},~\bfnm{Ao}\binits{A.}} \betal{et~al.}
(\byear{2001}).
\btitle{Analysis of gene expression identifies candidate markers and
pharmacological targets inprostate cancer}.
\bjournal{Cancer Res.}
\bvolume{61}
\bpages{5974--5978}.
\end{barticle}
%
\bptok{imsref}%
\endbibitem

%b63 ###
\bibitem[\protect\citeauthoryear{Wilcoxon}{1945}]{Wilcoxon45}
%
\begin{barticle}[author]
\bauthor{\bsnm{Wilcoxon},~\bfnm{F.}\binits{F.}}
(\byear{1945}).
\btitle{Individual comparisons by ranking methods}.
\bjournal{Biometrics}
\bpages{80--83}.
\end{barticle}
%
\bptok{imsref}%
\endbibitem

%b64 ###
\bibitem[\protect\citeauthoryear{Winslow et~al.}{2012}]{EmergCompBio2013}
%
\begin{barticle}[author]
\bauthor{\bsnm{Winslow},~\bfnm{Raymond}\binits{R.}},
\bauthor{\bsnm{Trayanova},~\bfnm{Nataliya}\binits{N.}},
\bauthor{\bsnm{Geman},~\bfnm{Donald}\binits{D.}} \AND
\bauthor{\bsnm{Miller},~\bfnm{Michael}\binits{M.}}
(\byear{2012}).
\btitle{The emerging discipline of computational medicine}.
\bjournal{Sci. Transl. Med.}
\bvolume{4}
\bpages{158rv11}.
\end{barticle}
%
\bptok{imsref}%
\endbibitem

%b65 ###
\bibitem[\protect\citeauthoryear{Xu, Geman and Winslow}{2007}]{TSPGPaper07}
%
\begin{barticle}[pbm]
\bauthor{\bsnm{Xu},~\bfnm{Lei}\binits{L.}},
\bauthor{\bsnm{Geman},~\bfnm{Donald}\binits{D.}} \AND
\bauthor{\bsnm{Winslow},~\bfnm{Raimond~L.}\binits{R.~L.}}
(\byear{2007}).
\btitle{Large-scale integration of cancer microarray data identifies a
robust common cancer signature}.
\bjournal{BMC Bioinformatics}
\bvolume{8}
\bpages{275}.
\bid{doi={10.1186/1471-2105-8-275}, issn={1471-2105},
pii={1471-2105-8-275}, pmcid={1950528}, pmid={17663766}}
\end{barticle}
%
\bptok{imsref}%
% NOT OUTPUTED:
% issn = 1471-2105
% fjournal = BMC bioinformatics
\endbibitem

%b66 ###
\bibitem[\protect\citeauthoryear{Xu et~al.}{2005}]{uTSPPaper}
%
\begin{barticle}[author]
\bauthor{\bsnm{Xu},~\bfnm{Lei}\binits{L.}},
\bauthor{\bsnm{Tan},~\bfnm{Aik~Choon}\binits{A.~C.}},
\bauthor{\bsnm{Naiman},~\bfnm{Daniel~Q.}\binits{D.~Q.}} \betal{et~al.}
(\byear{2005}).
\btitle{Robust prostate cancer marker genes emerge from direct
integration of inter-study microarray data}.
\bjournal{BMC Bioinformatics}
\bvolume{21}
\bpages{3905--3911}.
\end{barticle}
%
\bptok{imsref}%
\endbibitem

%b67 ###
\bibitem[\protect\citeauthoryear{Yao et~al.}{2004}]{ProstateMGDataSet}
%
\begin{barticle}[author]
\bauthor{\bsnm{Yao},~\bfnm{Za}\binits{Z.}},
\bauthor{\bsnm{Jaeger},~\bfnm{Jo}\binits{J.}},
\bauthor{\bsnm{Ruzzo},~\bfnm{Wo~L.}\binits{W.~L.}} \betal{et~al.}
(\byear{2004}).
\btitle{Gene expression alterations in prostate cancer predicting
tumor aggression and preceding development of malignancy}.
\bjournal{J.~Cli. Oncol.}
\bvolume{22}
\bpages{2790--2799}.
\end{barticle}
%
\bptok{imsref}%
\endbibitem

%b68 ###
\bibitem[\protect\citeauthoryear{Yao et~al.}{2007}]{MarfanDataSet}
%
\begin{barticle}[author]
\bauthor{\bsnm{Yao},~\bfnm{Zo}\binits{Z.}},
\bauthor{\bsnm{Jaeger},~\bfnm{Jo}\binits{J.}},
\bauthor{\bsnm{Ruzzo},~\bfnm{Wo}\binits{W.}} \betal{et~al.}
(\byear{2007}).
\btitle{A Marfan syndrome gene expression phenotype in cultured skin
fibroblasts}.
\bjournal{BMC Genomics}
\bvolume{8}
\bpages{319}.
\end{barticle}
%
\bptok{imsref}%
\endbibitem

%b69 ###
\bibitem[\protect\citeauthoryear{Yeang et~al.}{2001}]{SVM2}
\begin{barticle}[author]
\bauthor{\bsnm{Yeang},~\bfnm{Ce}\binits{C.}},
\bauthor{\bsnm{Ramaswamy},~\bfnm{Se}\binits{S.}},
\bauthor{\bsnm{Tamayo},~\bfnm{Peter}\binits{P.}} \betal{et~al.}
(\byear{2001}).
\btitle{Molecular classification of multiple tumor types.}
\bjournal{Bioinformatics}
\bvolume{17}
\bpages{S316--S322}.
\end{barticle}
%
%%
%(\byear{2001}).
%%
\bptok{imsref}%
\endbibitem

%b70 ###
\bibitem[\protect\citeauthoryear{Zhang, Yu and Singer}{2003}]{DT2}
%
\begin{barticle}[author]
\bauthor{\bsnm{Zhang},~\bfnm{Hu}\binits{H.}},
\bauthor{\bsnm{Yu},~\bfnm{Co~Y.}\binits{C.~Y.}} \AND
\bauthor{\bsnm{Singer},~\bfnm{Bo}\binits{B.}}
(\byear{2003}).
\btitle{Cell and tumor classification using}.
\bjournal{Proc. Natl. Acad. Sci. USA}
\bvolume{100}
\bpages{4168--4172}.
\end{barticle}
%
\bptok{imsref}%
\endbibitem

%b71 ###
\bibitem[\protect\citeauthoryear{Zhang et~al.}{2006}]{SVMNonConv}
%
\begin{barticle}[pbm]
\bauthor{\bsnm{Zhang},~\bfnm{Hao~Helen}\binits{H.~H.}},
\bauthor{\bsnm{Ahn},~\bfnm{Jeongyoun}\binits{J.}},
\bauthor{\bsnm{Lin},~\bfnm{Xiaodong}\binits{X.}} \AND
\bauthor{\bsnm{Park},~\bfnm{Cheolwoo}\binits{C.}}
(\byear{2006}).
\btitle{Gene selection using support vector machines with nonconvex penalty}.
\bjournal{Bioinformatics}
\bvolume{22}
\bpages{88--95}.
\bid{doi={10.1093/bioinformatics/bti736}, issn={1367-4803},
pii={bti736}, pmid={16249260}}
\end{barticle}
%
\bptok{imsref}%
% NOT OUTPUTED:
% issn = 1367-4803
% number = 1
% fjournal = Bioinformatics (Oxford, England)
\endbibitem

%b72 ###
\bibitem[\protect\citeauthoryear{Zhao, Logothetis and
Gorlov}{2010}]{ProstateTSP}
%
\begin{barticle}[author]
\bauthor{\bsnm{Zhao},~\bfnm{Hu}\binits{H.}},
\bauthor{\bsnm{Logothetis},~\bfnm{Co~J.}\binits{C.~J.}} \AND
\bauthor{\bsnm{Gorlov},~\bfnm{Ian~P.}\binits{I.~P.}}
(\byear{2010}).
\btitle{Usefulness of the top-scoring pairs of genes for prediction of
prostate cancer progression}.
\bjournal{Prostate Cancer Prostatic Dis.}
\bvolume{13}
\bpages{252--259}.
\end{barticle}
%
\bptok{imsref}%
\endbibitem

\end{thebibliography}
\end{document}